\documentclass[12pt]{article}
\usepackage{latexsym}
\usepackage{epsfig,amssymb,euscript,slashed}
\usepackage{amsmath}
\usepackage{cite} 
\usepackage{array,calc,epsfig}
\usepackage{bbm}
\usepackage{fancybox}
\usepackage[linktocpage]{hyperref}

\def\be{\begin{equation}}
\def\ee{\end{equation}}
\def\bseq{\begin{subequations}}
\def\eseq{\end{subequations}}

\def\bea{\begin{eqnarray}}
\def\eea{\end{eqnarray}}

\def\bseq{\begin{subequations}}
\def\eseq{\end{subequations}}
\def\beq{\begin{equation}}
\def\eeq{\end{equation}}

\numberwithin{equation}{section} %%
\usepackage[]{graphicx}

\def\ii {{\rm i}}

\def\sqr#1#2{{\vcenter{\vbox{\hrule height.#2pt
 \hbox{\vrule width.#2pt height#1pt \kern#1pt \vrule width.#2pt}\hrule
 height.#2pt}}}}

\def\slashchar#1{\setbox0=\hbox{$#1$}           % set a box for #1
\dimen0=\wd0                                 % and get its size
\setbox1=\hbox{/} \dimen1=\wd1               % get siste of /
\ifdim\dimen0>\dimen1                        % #1 is bigger
\rlap{\hbox to \dimen0{\hfil/\hfil}}      % so center / in box
#1                                        % and print #1
\else                                        % / is bigger
\rlap{\hbox to \dimen1{\hfil$#1$\hfil}}   % so center #1
/                                         % and print /
\fi}

\begin{document}
\font\cmss=cmss10 \font\cmsss=cmss10 at 7pt

\begin{flushright}{\scriptsize DFPD-16-TH-09 \\  
\scriptsize QMUL-PH-16-12}
\end{flushright}
\hfill
\vspace{18pt}
\begin{center}
{\Large 
\textbf{Correlators at large $c$ without information loss}
}
\end{center}

\vspace{8pt}
\begin{center}
{\textsl{Andrea Galliani$^{\,a, b}$, Stefano Giusto$^{\,a, b}$, Emanuele Moscato$^{\,c}$ and Rodolfo Russo$^{\,c}$}}

\vspace{1cm}

\textit{\small ${}^a$ Dipartimento di Fisica ed Astronomia ``Galileo Galilei",  Universit\`a di Padova,\\Via Marzolo 8, 35131 Padova, Italy} \\  \vspace{6pt}

\textit{\small ${}^b$ I.N.F.N. Sezione di Padova,
Via Marzolo 8, 35131 Padova, Italy}\\
\vspace{6pt}

\textit{\small ${}^c$ Centre for Research in String Theory, School of Physics and Astronomy\\
Queen Mary University of London,
Mile End Road, London, E1 4NS,
United Kingdom}\\
\vspace{6pt}

\end{center}

\vspace{12pt}

\begin{center}
\textbf{Abstract}
\end{center}

\vspace{4pt} {\small
\noindent 
We study a simple class of correlators with two heavy and two light operators both in the D1D5 CFT and in the dual AdS$_3 \times S^3 \times T^4$ description. On the CFT side we focus on the free orbifold point and discuss how these correlators decompose in terms of conformal blocks, showing that they are determined by protected quantities. On the gravity side, the heavy states are described by regular, asymptotically AdS$_3 \times S^3 \times T^4$ solutions and the correlators are obtained by studying the wave equation in these backgrounds. We find that the CFT and the gravity results agree and that, even in the large central charge limit, these correlators do not have (Euclidean) spurious singularities. We suggest that this is indeed a general feature of the heavy-light correlators in unitary CFTs, which can be relevant for understanding how information is encoded in black hole microstates.}

\vspace{1cm}

%\noindent {\em Possible comment ..........................................................................................................................................}

\thispagestyle{empty}

\vfill
\vskip 5.mm
\hrule width 5.cm
\vskip 2.mm
{
\noindent  {\scriptsize e-mails:  {\tt andrea.galliani@pd.infn.it, stefano.giusto@pd.infn.it, e.moscato@qmul.ac.uk, r.russo@qmul.ac.uk} }
}

\setcounter{footnote}{0}
\setcounter{page}{0}

\newpage

%%%%%%%%%%%%%%%%%%%%%%%%%%%%%%%%%%%%%%%%%
\section{Introduction}

From the early day of the AdS/CFT duality it has been conjuctured that type IIB string theory on AdS$_3\times S^3 \times {\cal M}$ (where ${\cal M}$ is either $T^4$ or $K3$) is dual to a $(4,4)$ 2D super Conformal Field Theory (CFT)~\cite{Maldacena:1997re}. Over the past 20 years this duality has been under intense scrutiny as it provides the ideal setup to study the microstates accounting for the entropy of a $1/8$-BPS large black hole~\cite{Strominger:1996sh}. A key question in this field is to what extent each individual black hole microstate is well described by the classical black hole solution when the theory is studied in a regime where supergravity is a good approximation. Since we are looking for possible breakdowns of the standard general relativity intuition, it seems a good approach to address this question by using the CFT description. Of course to make contact with the gravitational description, one needs to take the central charge $c$ to be large and focus on a particular (strongly interacting) point of the CFT moduli space. As we will see, it is however convenient to calculate many protected quantities at a different point where the CFT is described by a free orbifold~\cite{Dijkgraaf:1998gf,Larsen:1999uk,Jevicki:1998bm}.

On the CFT side, the black hole microstates correspond to $1/8$-BPS ``heavy'' states in the Ramond-Ramond sector which have conformal dimension of order $c$. A natural way to probe these states is to calculate the holographic 1-point functions of the different chiral primary operators. The general formalism was discussed in~\cite{Skenderis:2006uy,Skenderis:2006di} and was subsequently applied to the case of interest here first in the $1/4$-BPS sector~\cite{Skenderis:2006ah,Kanitscheider:2006zf,Kanitscheider:2007wq} and then for a class of $1/8$-BPS states~\cite{Giusto:2015dfa}. Another observable that can be used to study the heavy states is the entanglement entropy of an interval, since it is a well-defined CFT quantity~\cite{Calabrese:2004eu} that captures the geometric properties of the dual space-time~\cite{Ryu:2006bv,Hubeny:2007xt}\footnote{See~\cite{Asplund:2011cq,Bhattacharya:2012mi,Asplund:2014coa,Caputa:2014eta,Jones:2016iwx,David:2016pzn} for a general discussion of the entanglement entropy in heavy states.}. In the small interval limit~\cite{Calabrese:2010he} this entanglement entropy can be calculated for general heavy states of the $(4,4)$ CFT mentioned above and the result can be matched with the ones obtained from the dual geometries~\cite{Giusto:2014aba,Giusto:2015dfa}. Indeed higher order terms in the small interval expansion of the entanglement entropy are captured by the vev's of operators of increasing dimension (this connection has been recently emphasised also in~\cite{Beach:2016ocq}). From a technical point of view it is difficult to generalize the computation of 1-point functions to high dimensions, and hence this analysis only allows one to access the region of the microstate geometry close to the AdS boundary.  

In this paper we extend the study of the $1/4$ and $1/8$-BPS states in the $(4,4)$ CFT and their dual asymptotically AdS$_3\times S^3 \times {\cal M}$ geometries by studying the correlators of (two) light operators in a heavy state\footnote{Correlators involving two light twist operators that induce a transition between two different heavy states have been computed in~\cite{Lunin:2012gz}, with the purpose of studying absorption and emission of quanta from a D1D5 bound state.  A computation of two-point functions of primary operators in heavy excited states at large $c$ has also been recently performed in \cite{Goto:2016wme}.}. In the OPE limit in which the light operators are close, the correlator effectively resums an infinite series of vev's, and hence it represents an observable that can probe the bulk of the space-time. Our approach is based on very standard techniques: on the CFT side we need to calculate a 4-point function with two heavy and two light operators, while on the bulk side we study the wave equation of a light field in the dual non-trivial geometry. The main goal is to understand in some detail how the large $c$ limit of the CFT correlator reproduces the result obtained in the gravitational description. This heavy-light, large $c$ limit has been analyzed in several papers: an explicit expression for the Virasoro blocks in this limit was derived in~\cite{Fitzpatrick:2014vua,Fitzpatrick:2015zha} and a dual interpretation of this result in pure AdS$_3$ gravity was discussed in~\cite{Hijano:2015rla,Alkalaev:2015wia,Hijano:2015qja,Hijano:2015zsa,daCunha:2016crm}. Here we will apply the same approach to the simplest possible heavy operators in the $(4,4)$ CFT that have a dual geometric description in type IIB supergravity.

One of the main features of our analysis is that the full higher dimensional geometry is important in the bulk calculation, which is reflected on the CFT side by the contribution of operators that are not Virasoro descendants of the identity. This is a pattern that already emerged in the previous study of the 1-point functions and the entanglement entropy~\cite{Giusto:2014aba,Giusto:2015dfa} and, of course, it is particularly evident in our calculations because we chose very peculiar and simple heavy operators (i.e. very atypical states in the black hole ensemble). However, these examples show that pure heavy states are not directly described by the 3D geometry of the BTZ solution and that, on the CFT side, Virasoro primaries different from the identity can play an important role also in the large $c$ limit. In particular, in the correlators we consider, the singularities due to the large $c$ Virasoro block of the identity are resolved by the contributions of new primaries that are non-trivial already at the leading order in the limit $c\gg 1$. So in this case the pattern is different from the one discussed in~\cite{Fitzpatrick:2016ive}\footnote{See~\cite{Perlmutter:2015iya,Beccaria:2015shq,Fitzpatrick:2015dlt} for a detailed discussion of the Virasoro blocks beyond the leading term in the $c\to\infty$ expansion and \cite{Anous:2016kss} for the possible relevance of $1/c$ corrections in black hole collapse.}, where it is argued that $1/c$ corrections are crucial to restore unitarity. In the simple cases we investigate, this mechanism is visible already at the supergravity level as the relevant new Virasoro primaries are actually affine descendants of the identity. For more general correlators the contribution of primary operators that are not captured in the supergravity approximation will most likely be crucial to avoid the appearance of spurious singularities when $c\to\infty$. In the conclusions we present an argument based on crossing symmetry supporting the idea that the heavy-light correlators have in general a regular large $c$ limit if the contribution of all primaries is considered. Thus, even if the results for the correlators we studied cannot be directly extrapolated to typical black hole microstates, we suggest that the absence of large $c$ spurious singularities in the heavy-light correlators is generic and that it might be seen as a CFT feature supporting the ``fuzzball'' proposal~\cite{Mathur:2005zp,Mathur:2008nj}.

We conclude the introduction with an outline of the structure of the paper. In section~\ref{sec:cfts} we describe the $(4,4)$ CFT of interest in terms of the symmetric orbifold ${\cal M}^N/S_N$. Even if this is appropriate for a point in the moduli space that is far from the regime where supergravity is valid, the free orbifold description provides a simple way to characterize the operators we use and calculate the correlators we are interested in. In section~\ref{sec:bd} we analyse the same correlators in terms of Virasoro blocks in order to highlight that non-trivial primaries contribute also in the large $c$ limit. For the examples under analysis, it turns out that these new Virasoro primaries are actually part of the identity affine block of the R-symmetry and, in particular, the full answer is captured by the $U(1)$ affine blocks. This shows that the correlators we focus on are fully determined by protected quantities and so it should be possible to reproduce the same results by a gravity calculation. This is discussed in section~\ref{sec:grav} where the geometries dual to the heavy states are introduced. Then we extract holographically the 4-point correlators discussed on the CFT side and show that the two results match. In section~\ref{sec:discussion} we discuss the possible generalization of our analysis and also its possible relevance for the problem of characterising the microstates of the Strominger-Vafa black hole. The appendices~\ref{sec:cftconv} and~\ref{sec:waveequation} contain a summary of the technical results useful for the CFT the gravity analysis respectively.

%%%%%%%%%%%%%%%%%%%%%%%%%%%%%%%%%%%%%%%%%
\section{The CFT picture}
\label{sec:cfts}

In this section we discuss some simple examples of four-point correlators in the D1D5 CFT. In particular we are interested in correlators with two heavy ($O_H$) operators, which have conformal dimension of order $c$, and two light ($O_L$) operators, which have conformal dimension of order one. Thus the structure of the correlators we consider is
  \begin{equation}
\label{eq:cHHLL}
  \langle O_H(z_1) \bar{O}_H(z_2) O_L(z_3) \bar{O}_L(z_4) \rangle = \frac{1}{z_{12}^{2h_H} z_{34}^{2h_L}} \frac{1}{\bar{z}_{12}^{2\bar{h}_H} \bar{z}_{34}^{2\bar{h}_L}} 
{\cal G}(z,\bar{z})  ~,~~
 \end{equation}
where, as usual, $z_{ij}=z_i-z_j$ and 
\begin{equation}
 \label{eq:zdef}
z=\frac{z_{14} z_{23}}{z_{13}z_{24}}\,,
\end{equation}
while $(h_H,\bar{h}_H)$ and $(h_L,\bar{h}_L)$ are the holomorphic/anti-holomorphic conformal dimensions of the heavy and light operators respectively.

We will take two main simplifying assumptions. First we focus on highly supersymmetric operators. The light operators we use are chiral primaries both in the left and in the right sector of the CFT. Instead the heavy operators are in the Ramond-Ramond sector of the CFT, but are related to chiral primaries by a chiral algebra transformation that acts only on the left sector (hence they generically preserve half of the CFT supercharges).  Second we work at the free orbifold point of the CFT moduli space, where the theory, which has central charge $c= 6 N$, is described by a collection of $N$ copies of free fields (we also call each such copy a ``strand'' of length $1$). In each copy we have four free bosons and four free fermions which are labelled by the indices $(\alpha,\dot{\alpha})$ of the $SU(2)_L \times SU(2)_R$ R-symmetry group and the indices $(A,\dot{A})$ of the $SU(2)_1 \times SU(2)_2$ group acting on the tanget space of $T^4$ or $K3$. Thus the collection of elementary fields we use is\footnote{We will use the conventions of~\cite{Giusto:2015dfa}, which are based on~\cite{Avery:2010qw}. See appendix~\ref{sec:cftconv} for more details.}
\begin{equation}
  \label{eq:Xpsi}
  \left(X^{A \dot{A}}_{(r)}(\tau,\sigma)\,,\;\psi^{\alpha \dot{A}}_{(r)}(\tau+\sigma)\,,\;\tilde\psi^{\dot\alpha \dot{A}}_{(r)}(\tau-\sigma)\right)\;,
\end{equation}
where $r=1,\ldots,N$ labels the copy and $(\tau,\sigma)$ are the standard coordinates on the CFT space-time.

As a final comment, if one sees $K3$ as a $Z_2$ orbifold of $T^4$, all our operators are in the untwisted sector of the $Z_2$-orbifold action, which multiplies by $-1$ the operators with an odd number of indices $\dot{A}$. Thus the correlators we consider are relevant both for the $(T^4)^N/S_N$ and $(K3)^N/S_N$ CFTs.

\subsection{Simple correlators in the untwisted sector}

We first focus on operators in the untwisted sector of the symmetric orbifold, which means that they are written as combinations of operators acting on each copy.  The symmetry under permutations among the copies is realised differently in the light and the heavy operators: the light operators act trivially on all the strands but one, while the heavy ones are constructed by multiplying $N$ copies of the same operator, each copy acting on a different strand:
 \begin{equation}
  \label{eq:HLO}
  O_L = \frac{1}{\sqrt{N}}\sum_{r=1}^N O^L_{(r)}~,~~   
  O_H = \otimes_{r=1}^N O^H_{(r)}~.
  \end{equation}
In this article we concentrate on light operators of dimension $h_L=\bar h_L=1/2$ constructed with the fermions; in concrete we take
\begin{equation}
  \label{eq:Opp}
O^L_{(r)} = -\frac{\ii}{\sqrt{2}} \psi^{1\dot{A}}_{(r)} \epsilon_{\dot{A} \dot{B}} \tilde{\psi}^{\dot{1} \dot{B}}_{(r)} \equiv O^{++}_{(r)}~.
\end{equation}
All the operators $O^H_{(r)}$ we are going to consider in the untwisted sector have right conformal dimension $\bar{h}_{(r)} = 1/4$ and right spin $\bar{j}_{(r)} =
1/2$, which gives a total right conformal dimension for the heavy operators $\bar{h}_H = N/4$, so we can distinguish the heavy operators by their left conformal dimension and left spin. The heavy operators we choose in the untwisted sector are characterised by an integer $s$ determining the number of $J^+$ excitations acting on a ground state in each copy; their explicit expression is more easily written in the bosonized language (see \eqref{eq:Spines}), and their left conformal dimension and spin are given by
\begin{equation}
 \label{eq:hjleftuntwist}
h_H = N \left( s + \frac{1}{2} \right)^2,\qquad j_H = N \left (s+\frac{1}{2} \right).
\end{equation}
We therefore denote the single copy operators making up the heavy states as $O^H_{(r)}(s)$ and the same notation will be adopted for the correlators, which are denoted as $\mathcal{G}\left(s; z, \bar{z}\right)$.

As a first concrete example we consider the heavy operator corresponding to $s = 0$; it is  written in terms of the spin fields $S^{\dot{A}}_{(r)}$  twisting the elementary fermions $\psi_{(r)}^{\alpha \dot{A}}$ (and $\tilde{S}^{\dot{A}}_{(r)}$  twisting  $\tilde{\psi}_{(r)}^{\dot{\alpha} \dot{A}}$)
\begin{equation}
  \label{eq:OLM}
  O^H_{(r)}(s=0) = S^{\dot{1}}_{(r)} S^{\dot{2}}_{(r)} \tilde{S}^{\dot{1}}_{(r)} \tilde{S}^{\dot{2}}_{(r)}~.
\end{equation}
Let us comment on the AdS-dual interpretation of the operators entering this correlator. The heavy state is the Ramond-Ramond ground state with the highest value for the left ($J^3_0$) and right ($\tilde{J}^3_0$) spins allowed by unitarity. This state can be obtained by starting from the $SL(2,\mathbb{C})$ invariant vacuum and performing a spectral flow to the Ramond-Ramond sector, which means that the dual supergravity solution\footnote{It is possible to extend this solution to an asymptotically flat type IIB supergravity background, which then represents a (very special) microstate for the Strominger-Vafa black hole~\cite{Balasubramanian:2000rt,Maldacena:2000dr}.} is locally isometric to AdS$_3\times S^3$. The light operator \eqref{eq:Opp} is a supersymmetric fluctuation of the $B$-field  and the axion~\cite{Kanitscheider:2007wq} around the geometry dual to $O_H$. We can calculate the correlator at the orbifold point of the CFT moduli space by using the standard bosonization approach and the free field contractions in the bosonic language. We collect in Appendix~\ref{sec:cftconv} our conventions and a brief derivation of the result:
\begin{equation}
  \label{eq:LMOO}
  {\cal G}\bigl(s=0; z, \bar{z}\bigr) = \frac{1}{|z|}~.
\end{equation}

A simple generalization of~\eqref{eq:LMOO} is to consider the correlator with the same light states, but heavy states corresponding to generic $s$, which contain excited spin fields in the holomorphic sector
\begin{equation}
  \label{eq:OLMs}
  O^H_{(r)}\left( s; z, \bar{z}\right) = S^{\dot{1}}_{s (r)} S^{\dot{2}}_{s (r)} \tilde{S}^{\dot{1}}_{(r)} \tilde{S}^{\dot{2}}_{(r)}~,
\end{equation}
where $S^{\dot{A}}_{s (r)}$ has conformal weight $(s+1/2)^2/2$. Again by using the bosonized language it is straightforward to calculate the correlator (see Appendix~\ref{sec:cftconv} for some detail)
\begin{equation}
  \label{eq:LMOOs}
  {\cal G}\left(s;z,\bar{z}\right) = \frac{1}{z^{s+\frac{1}{2}} \bar{z}^{\frac{1}{2}}}
\end{equation}
Note that the new heavy state is an affine descendant of the Ramond-Ramond ground state~\eqref{eq:OLM} and so the dual description can be locally mapped to AdS$_3\times S^3$ with a change of coordinates that encode (at the boundary) the action of the superalgebra on~\eqref{eq:OLM}. Thus, as discussed later in Section~\ref{section:twisted}, this new correlator inherits several properties from the previous example in~\eqref{eq:LMOO}. 

\subsection{Simple correlators in the twisted sector}
 \label{section:twisted}
We now consider correlators in the twisted sector of the CFT, meaning that the $N$ copies are divided into $N/k$ bunches, each made of $k$ copies glued together. We call each bunch a strand of length $k$. Within a strand of length $k$, we have $k$ elementary bosons and fermions with non-trivial periodicities, as written for instance for the $\psi^{\alpha \dot{A}}_{(r)}$ in~\eqref{eq:fermionlengthk}. As usual we can take the linear combinations~\eqref{eq:fermionsrho} and define new fields which diagonalize the boundary conditions. We label these twisted sectors with $\rho$: for instance the fermions $\psi^{\alpha\dot{A}}_\rho$, with $\rho = 0,\ldots,k-1$, have the standard twisted boundary conditions~\eqref{eq:diagonalizedperiod}. In analogy to what we did in the previous section, the heavy operators are constructed by taking $N/k$ identical strands of length $k$. The anti-holomorphic conformal dimension of our heavy operators on each strand is always $\bar{h}_H=k/4$ and their right spin is $\bar{j}_H = 1/2$. As before, we consider $s$ momentum-carrying excitations in the holomorphic sector, so we characterize the heavy operators by two integers $s$ and $k$, and their left conformal dimension and spin read
\begin{equation}
 \label{eq:hjlefttwist}
h_H = \frac{N}{k} \left(\frac{k}{4}+\frac{s(s+1)}{k}\right),\qquad j = \frac{N}{k} \left(s+\frac{1}{2}\right)\,.
\end{equation}
The operators are denoted as $O_H(s, k)$ and the correlators as $\mathcal{G}(s, k; z, \bar{z})$.

The first kind of heavy operators we consider corresponds to $s=0$ and generic $k$ and is a generalization to strands of length $k$ of \eqref{eq:OLM}: on each strand we have $k$ operators $S^{\dot{A}}_{k,\rho}$ and $k$ operators $\tilde{S}^{\dot{A}}_{k,\rho}$ and the total heavy operator is
\begin{equation}
O_H\!\left( s=0, k\right) = \left[\Sigma_k\, \tilde{\Sigma}_k \otimes_{\rho=0}^{k-1}\ S^{\dot{1}}_{k,\rho} S^{\dot{2}}_{k,\rho} \tilde{S}^{\dot{1}}_{k,\rho} \tilde{S}^{\dot{2}}_{k,\rho}\right]^{N/k},
\end{equation}
where $\Sigma_k$ ($\tilde{\Sigma}_k$) is the twist field inducing on the bosonic fields $\partial X^{A\dot{A}}$ ($\bar\partial X^{A\dot{A}}$) the same identification specified in the fermionic sector by~\eqref{eq:fermionlengthk}. The correlator is obtained again through bosonization in the twisted sector (the derivation is sketched in appendix~\ref{sec:cftconv}) and reads
\begin{align}
\mathcal{G}\left(s= 0,k; z, \bar{z}\right) = \frac{1/k}{|z|}\ \frac{1-|z|^2}{1-|z|^{2/k}},
\end{align}
where the $1/k$ factor comes from having the same contribution from each of the $N/k$ strands and from the normalization chosen for the light operators in \eqref{eq:HLO}.

The second kind of heavy operator we consider corresponds to nonzero $s$ and $k$ and is a generalization to strands of length $k$ of \eqref{eq:OLMs}. These states have $s(s+1)/k$ units of momentum on each strand, and since the number of momentum units must be integer, assuming $k$ is a prime number for simplicity, we have that either $s = pk$ or $s=pk-1$, with $p\in \mathbb{N}$. In the $s=pk$ case, in the left sector of each strand we have $k$ operators $S^{\dot{A}}_{k,s,\rho}$, and another $k$ operators $\tilde{S}^{\dot{A}}_{k,\rho}$ live in the right sector. The total heavy operator is
\begin{equation}
O_H\!\left( s=pk,k\right) = \left[ \Sigma_k \,\tilde{\Sigma}_k \otimes_{\rho=0}^{k-1}\ S^{\dot{1}}_{k,pk,\rho} S^{\dot{2}}_{k,pk,\rho} \tilde{S}^{\dot{1}}_{k,\rho} \tilde{S}^{\dot{2}}_{k,\rho}\right]^{N/k}.
\end{equation}
Notice that since $h_H$ depends on $s$, for $s>0$ we have $h_H \neq \bar{h}_H$ and so heavy states carry non-vanishing momentum; the correlator reads
\begin{equation}
 \label{eq:resultsfk}
\mathcal{G} \left( s=pk,k; z, \bar{z}\right) =  \frac{1/k}{|z|}\ \frac{1-|z|^2}{1-|z|^{2/k}}\,z^{-p}.
\end{equation}
When $s=pk-1$ the heavy operator differs from the previous case only in the $\rho=0$ sector, and has the form
\begin{equation}
 \label{eq:heavyk-1}
O_H(s=pk-1,k) = \left[\Sigma_k \tilde{\Sigma}_k \hat{S}^{\dot{1}}_{k,0} \hat{S}^{\dot{2}}_{s,0} \otimes_{\rho=1}^{k-1} S^{\dot{1}}_{k,pk,\rho} S^{\dot{2}}_{k,pk,\rho} \tilde{S}^{\dot{1}}_{k,\rho} \tilde{S}^{\dot{2}}_{k,\rho}\right]^{N/k},
\end{equation}
and the correlator reads
\begin{equation}
 \label{eq:resultsfk-1}
\mathcal{G}(s=kp-1,k;z, \bar{z}) = \frac{1/k}{|z|}\,z^{-p}\,\left( z + \frac{|z|^{2/k} -|z|^2}{1-|z|^{2/k}} \right).
\end{equation}

%%%%%%%%%%%%%%%%%%%%%%%%%%%%%%%%%%%%%%%%%

\section{Conformal blocks decomposition}
\label{sec:bd}

In this section we analyze the correlators obtained above in terms of Virasoro and affine conformal blocks, exploiting the underlying $SU(2)$ R-symmetry.  In the channel where the two light operators approach each other ($z_3\to z_4$), the cross-ratio $z$ tends to 1 and we can expand the function ${\cal G}$ in~\eqref{eq:cHHLL} to extract the Virasoro or affine primary operators entering in the decomposition:
\begin{equation}
  \label{eq:bd}
   {\cal G} = (1-z)^{2h_L} (1-\bar{z})^{2\bar{h}_L} \sum_{O_p} C_{HHO_p} C_{LLO_p} {\cal V}_{V,A}(h_p,h_H,h_L,z)  \bar{\cal V}_{V,A}(\bar{h}_p,\bar{h}_H,\bar{h}_L,\bar{z})\;,
\end{equation}
where the sum is over all Virasoro or affine primaries $O_p$, ${\cal V}_{V}$ and ${\cal V}_{A}$ are the Virasoro or affine blocks, and $C_{HHO_p}$ ($C_{LLO_p}$) are the structure constants between $O_p$ and the heavy (light) operators. 

\subsection{Virasoro blocks decomposition}
\label{sec:vbd}

For the description in terms of the Virasoro blocks we focus on the large $c$ limit where it is possible to use the results of~\cite{Fitzpatrick:2014vua,Fitzpatrick:2015zha}. In this limit the contribution of the Virasoro descendents of a primary of weight $h_p$ is captured by the block whose holomorphic part is\footnote{We normalize the conformal block so that the first term of the $z\to 1$ expansion is $(1-z)^{h_p-2h_L}$.} 
\begin{equation}
  \label{eq:VBg}
  {\cal V}_V(h_p, h_H, h_L,z) = z^{h_L\left(\alpha-1\right)} \left(\frac{1-z^\alpha}{\alpha} \right)^{h_p- 2 h_L}\, {}_2F_1\left(h_p,h_p;2 h_p;1-z^\alpha\right)~,
\end{equation}
where $\alpha= \sqrt{1-\frac{24 h_H}{c}}$. Some of the heavy states we consider have conformal dimension $h_H=c/24$; in this case the large $c$ limit of the Virasoro block is captured by the $\alpha\to 0$ limit\footnote{It is also possible to follow a similar derivation as in~\cite{Fitzpatrick:2015zha} with $h_H=c/24$ and show that the result agrees with the $\alpha\to 0$ limit of the formula above.} of~\eqref{eq:VBg}
\begin{equation}
  \label{eq:VBgc24}
  {\cal V}_V(h_p, h_H\to c/24, h_L,z) = z^{-h_L} \left(-{\ln z}\right)^{h_p-2 h_L}~.
\end{equation}

In all amplitudes analyzed in the previous section, the first primary entering the $z\to 1$ decomposition is the identity. If we consider only the contribution of its Virasoro block, for instance in the simplest case~\eqref{eq:LMOO}, we have
\begin{equation}
  \label{eq:LMOOib}
  {\cal G}\bigl(s=0; z, \bar{z}\bigr) = \frac{1}{|z|} \frac{|1-z|^2}{|\ln z|^2}+\ldots~,
\end{equation}
where we used~\eqref{eq:bd} and~\eqref{eq:VBgc24} with $h_p=0$, $h_L=1/2$, and the analogous expression for the anti-holomorphic sector with $\bar{h}_p=0$, $\bar{h}_L=1/2$. Focusing on the holomorphic dependence, there is a mismatch between~\eqref{eq:LMOOib} and~\eqref{eq:LMOO} already at the order $(1-z)$, which signals that primaries of conformal dimension $(h_p,\bar h_p)=(1,0)$ must contribute to the correlator~\eqref{eq:LMOO}. It is straightforward to see that in the OPE of the two light operators $O_L$, $\bar O_L$ the first (normalized) Virasoro primaries are 
\begin{equation}
  \label{eq:fvp}
\begin{aligned}
O_{(1,0)} &= \sqrt{\frac{2}{N}} \sum_{r=1}^N J^3_{(r)}\;,\\
O_{(2,0)} &= \frac{1}{\sqrt{6 N}} \sum_{r=1}^N \left( - \partial\psi^{\alpha\dot{A}}_{(r)}\psi^{\beta\dot{B}}_{(r)} \epsilon_{\alpha\beta} \epsilon_{\dot{A}\dot{B}}  + \frac{1}{2} \partial X_{(r)}^{A\dot{A}} \partial X_{(r)}^{B\dot{B}}\epsilon_{AB}\epsilon_{\dot{A}\dot{B}}\right)\;.
\end{aligned}
\end{equation}
We can straightforwardly compute the three-point correlators between these primaries and the heavy or the light operators so to extract the structure constants entering in the decomposition~\eqref{eq:VBg}. For later convenience, we summarize the results involving the light and the heavy operators in~\eqref{eq:OLMs} for generic $s$:
\begin{equation}
  \label{eq:Cc24}
  \begin{aligned}
C_{LLO_{(1,0)}} = \frac{1}{\sqrt{2}}\;,~~ &
C_{HHO_{(1,0)}} = \sqrt{2} \left(s + \frac{1}{2}\right)\;,~~\\
C_{LLO_{(2,0)}} = \frac{1}{\sqrt{6}}\;,~~ &  
C_{HHO_{(2,0)}} = \frac{(1+2 s)^2}{2 \sqrt{6}}\,.
  \end{aligned}
\end{equation}
Thus one can improve on the decomposition~\eqref{eq:LMOOib} by adding the Virasoro blocks for the operators in~\eqref{eq:fvp}
\begin{equation}
  \label{eq:LMOOijb}
  {\cal G}\bigl(s=0; z, \bar{z}\bigr) = \frac{1}{|z|} \frac{|1-z|^2}{|\ln z|^2} \left(1 - \frac{1}{2} \ln z + \frac{1}{12} (\ln z)^2 +\ldots \right)~,
\end{equation}
which reproduces~\eqref{eq:LMOO} to the leading order in the $\bar{z}\to 1$ and to second order in ${z}\to 1$ limits.  

We can proceed with the same analysis for the remaining correlator \eqref{eq:LMOOs} in the untwisted sector. One now has $h_H = \frac{c}{6}\left(s+\frac{1}{2}\right)^2, \bar{h}_H = \frac{c}{24}$, and we have to use the large $c$ Virasoro blocks~\eqref{eq:VBg} for the holomorphic part and \eqref{eq:VBgc24} for the anti-holmorphic one. The contribution of the identity gives
\begin{equation}
  \label{eq:SFuntwistedib}
  {\cal G}\bigl(s; z, \bar{z}\bigr) = - \frac{|1-z|^2}{\sqrt{\bar{z}} \log(\bar{z})}\ \frac{\alpha\ z^{\frac{\alpha-1}{2}}}{1-z^\alpha} + \cdots,
\end{equation}
where $\alpha = \sqrt{1 - 4\left(s+\frac{1}{2}\right)^2}$. Again, the expansion of the expression above for $z\to1$ already disagrees with the exact result \eqref{eq:LMOOs} at order $(1-z)$ and, as before, we need to add the Virasoro blocks of other primaries. By using the $s$-dependent structure constants in~\eqref{eq:Cc24}, we have 
\begin{align}
{\cal G}\bigl(s; z, \bar{z}\bigr) =  \frac{|1-z|^2}{\sqrt{\bar{z}} \log(\bar{z})}\ \frac{\alpha\ z^{\frac{\alpha-1}{2}}}{z^\alpha-1} \left[
1 - \frac{1+2s}{2} \log z - \frac{(1+2 s)^2}{2\alpha^2} \left(2 +\frac{1+z^\alpha}{1-z^\alpha} \log z^\alpha \right) +\ldots \right] .
\end{align}
As in the $s=0$ case, the expression above agrees with the exact result \eqref{eq:LMOOs} up to order $(1-z)^2 (1-\bar{z})^0$ in the $z\to1$ expansion.

\subsection{Affine blocks decomposition}
\label{sec:abd}

In all our examples the light operator~\eqref{eq:Opp} used to probe the heavy states is written just in terms of the elementary fermions of the orbifold CFT. This suggests that it is convenient to study the decomposition of this type of correlators in terms of affine blocks related to the $SU(2)_L$ current algebra~\eqref{eq:Jtot}. As this symmetry is part of the chiral superalgebra we can use this analysis to argue that the correlators considered in the previous section are protected by supersymmetry, and then, in the next section, to match the free CFT result with supergravity calculations. Also, in contrast to the pure Virasoro case, the results for the affine blocks are exact in $c$ and so we can use them to understand the effect of resumming the large $c$ limit of the blocks of all Virasoro primaries: we will see that the singularities due to each Virasoro block~\cite{Fitzpatrick:2016ive} disappear even at large $c$. This is reminiscent of what happens in some out-of-time-ordered correlators in $SU(N)_k$ WZW models~\cite{Caputa:2016tgt}.

We start from the simplest example discussed in~\eqref{eq:LMOO} and analyze it in two slightly different ways. First we observe that the correlator is purely fermionic and that it is given by a sum over the $N$ strands of correlators that involve non-trivially only the fields on one strand at a time. We can then effectively restrict to two free complex fermions on a length one strand, which realize a $SU(2)_{k=1} \times U(1)$ WZW model\footnote{This approach is similar to one adopted in~\cite{Caputa:2016tgt} in the study of quantum chaos in rational CFT. Notice however that in that analysis the large central charge limit is obtained by studying the WZW model $SU(N)_k$ in the limit $N,k\to\infty$ with $N/k$ fixed, instead of using the symmetric orbifold of many copies of $SU(2)_{k=1}$, as relevant for our case.} (see for instance~\cite{DiFrancesco:1997nk}). Note that the $SU(2)_{k=1}$ factor is identified with the R-symmetry $SU(2)_L$, and is thus a symmetry of the CFT at a generic point in the moduli space; the $U(1)$ symmetry, instead, disappears away from the free orbifold point. The non-trivial 4-point function to compute is the one appearing in the first line of~\eqref{eq:corrcompuntwist} for $s=0$;  with respect to the $SU(2)_{k=1}$ subsector of the WZW model, all the four operators involved are $SU(2)_{k=1}$ primaries of spin $1/2$. Though the light operators also carry a $U(1)$ charge, the heavy states are scalars under this $U(1)$, and thus the correlator reduces to a trivial 2-point function in the $U(1)$ sector. This means that it should be possible to write the amplitude~\eqref{eq:LMOO} by using the classic result of~\cite{Knizhnik:1984nr} for the affine blocks of $SU(N)_k$ WZW models in the special case where $N=2$ and $k=1$. This model has only two primaries (the identity and the spin $1/2$ primary) and so the only $SU(2)_{k=1}$ primary appearing in the OPE of two spin $1/2$ operators has to be the identity. So in this case the affine decomposition~\eqref{eq:bd} contains just one term, given by the $SU(2)_{k=1}$ block of the identity: since, as we said, $SU(2)_{k=1}$ is part of the superconformal algebra, this shows that the amplitude~\eqref{eq:LMOO} can be written in terms of protected quantities.

It is straightforward to check that the hypergeometric describing the $SU(N)_k$ blocks reduce to elementary functions for the identity block with $N=2$ and $k=1$; by adapting the results summarized in~\cite{DiFrancesco:1997nk} to our notations we have\footnote{In order to translate the choice of the $z_i^{D}$'s of~\cite{DiFrancesco:1997nk} into ours it is sufficient to take $z_{i=1,3}^{D}=z_{i=1,3}$, $z_2^{D}=z_4$, and $z_4^{D}=z_2$; notice also that the blocks in~\cite{DiFrancesco:1997nk} have a different normalization and that the hypergeometric appearing in~Eq.(15.170) of~\cite{DiFrancesco:1997nk} should read ${}_2F_1\left(\frac{\kappa+1}{\kappa},\frac{\kappa-1}{\kappa},\frac{2\kappa-N}{\kappa},x\right)$.}
\begin{equation}
  \label{eq:n2k1ab}
  {\cal V}_{SU(2)_1} = (1-z)^{-2 h_L}
  \left(\begin{array}{c}
    F^-_1 \\ F^-_2
  \end{array}\right) = (1-z)^{-2 h_L}
  \left(\begin{array}{c}
    z^{-\frac 12} \\ z^{\frac 12}
  \end{array}\right)\;,
\end{equation}
where the component $F^-_1$ ($F^-_2$) contributes if the operators in $z_1$ and $z_4$ ($z_2$ and $z_4$) have opposite spin. In our case~\eqref{eq:corrcompuntwist} $F^-_1$ enters in the decomposition of~\eqref{eq:LMOO} and reproduces directly the whole amplitude.

The simple result in~\eqref{eq:n2k1ab} suggests that only a subsector of the full $SU(2)_{k=1}$ affine blocks contributes to our correlator. This is indeed the case and the amplitude is saturated just considering the affine descendants obtained by acting with the modes of the currents $J^3$ (and $\tilde{J}^3$) on the identity. Focusing on this  $U(1)_L$ subgroup, the affine block of the identity reads\footnote{See~\cite{Fitzpatrick:2015zha} for a recent discussion of the $U(1)$ blocks in the context of the heavy-light large $c$ limit.}  
\begin{equation}
  \label{eq:abu1}
  {\cal V}_{U(1)}(q_H,q_L,z) = (1-z)^{-2 h_L} z^{2 q_H q_L}~,
\end{equation}
where the $q_H$ and $q_L$ are identified with the $J^3_0$ quantum numbers ($j$) of the operators $\bar{O}^H_{(r)}(z_2)$ and $O^L_{(r)}(z_3)$ (note that, with this identification, the level of the $U(1)_L$ current algebra is $k=1/2$, in the conventions of \cite{Fitzpatrick:2015zha}). Then, by using $q_H=-1/2-s$ and $q_L=1/2$, we immediately reproduce not just~\eqref{eq:LMOO} but also~\eqref{eq:LMOOs}.

The correlators involving states in the twisted sector can also be described in terms of $U(1)_L$ affine blocks. From~\eqref{eq:J3rho} the generator $J^3$ on a strand of length $k$ splits into the sum of $k$ $U(1)$'s labelled by $\rho=0,\ldots,k-1$. While the charge of the light operator is still $q_L=1/2$ for any $\rho$, the charge of the heavy operators is $\rho$-dependent, as can be seen from~\eqref{eq:heavyoptwist} and~\eqref{eq:Shatop}. So the contribution to the block decomposition of each $\rho$-sector is given by~\eqref{eq:abu1} with the values for the $q$'s that can be read off from \eqref{eq:heavyoptwist} and~\eqref{eq:Shatop}; after performing the sum over $\rho$, one can check that the correlators \eqref{eq:resultsfk} and \eqref{eq:resultsfk-1} are reproduced by~\eqref{eq:bd} with only the inclusion of the $U(1)_L$ affine block of the identity.

%%%%%%%%%%%%%%%%%%%%%%%%%%%%%%%%%%%%%%%%%

\section{The gravity picture}
\label{sec:grav}

Let $|s,k\rangle$ denote the pure states generated by the action of the heavy operators on the conformal invariant vacuum:
\begin{equation}
|s,k\rangle \equiv \lim_{z,\bar z\to0} O_H(s,k;z,\bar z) |0\rangle\,. 
\end{equation}
 Since operators of conformal dimension of order $c$ backreact strongly on the geometry and generate a non-trivial gravity background, these states admit a dual gravity description. The four-point correlators computed in the previous section can thus be thought as two-point functions of light correlators in a non-trivial geometry:
 \begin{equation}
 \langle s,k | O_L(1) {\bar O}_L(z) |s,k\rangle = \frac{1}{|1-z|^{4 h_L}}\,\mathcal{G}(z,\bar z)\,. 
 \end{equation}
 In the limit of large central charge this geometry is well approximated by a solution in supergravity. In this section we will compute this two-point function at the point in the CFT moduli space where supergravity is weakly coupled, i.e. higher curvature corrections are negligible. 

This point in moduli space differs from the free orbifold point, where the CFT correlators have been computed. While the light operators we consider are chiral primaries both in the left and right sector and the heavy operators are chiral at least in the right sector, their four-point correlators are generically expected to receive corrections when one deforms the free orbifold theory towards the point in moduli space corresponding to weakly coupled supergravity. This is made evident by the decomposition~\eqref{eq:bd}, which generically contains also non-chiral primaries (and their descendants). For the particular correlators we consider in this paper, we have however shown in Section~\ref{sec:abd} that the expansion \eqref{eq:bd} only contains the identity operator and its super-descendants with respect to a $U(1)$ subgroup of the superconformal algbera. This implies that CFT and gravity results must agree. In this section we verify this expectation.

\subsection{The 6D geometries}

The D1D5 CFT is dual to a gravity theory on spaces that are asymptotically\footnote{To describe generic states one should consider the full 10D geometry, which asymptotes AdS$_3\times S^3\times M$, with $M$ either $T^4$ or $K3$. For the class of states we consider, the $M$ factor is irrelevant and we restrict to the 6D part of the geometry.} AdS$_3\times S^3$: the $S^3$ factor is necessary to geometrically implement the $SU(2)_L\times SU(2)_R$ R-symmetry of the CFT. The geometries generated by generic heavy operators are complicated 6D spaces, which only asymptotically factorize into the product of AdS$_3$ and $S^3$. All these geometries are known when the heavy operators are chiral primaries both on the left and the right sector \cite{Lunin:2001jy,Lunin:2002iz,Kanitscheider:2007wq}; a subset of geometries is known for heavy operators that are chiral only on the right sector \cite{Giusto:2004id,Bena:2005va,Berglund:2005vb,Ford:2006yb,Lunin:2012gp,Giusto:2012yz,Bena:2015bea,Bena:2016agb}, or are not chiral on either sector \cite{Jejjala:2005yu,Chakrabarty:2015foa}. 

In this article we concentrate on a particularly simple set of BPS states, whose dual geometries are {\em locally} isometric to AdS$_3\times S^3$ via a diffeomorphism that does not vanish at the boundary. The 6D Einstein metric for these states can be written in the form
\begin{subequations}\label{eq:6Dmetric}
\begin{align}
ds^2&=\sqrt{Q_1 Q_5}\,(ds^2_{AdS_3}+ds^2_{S^3})\,,\\
ds^2_{AdS_3}&=\frac{d r^2}{a^2 k^{-2}+r^2}-\frac{a^2 k^{-2}+r^2}{Q_1Q_5}dt^2+\frac{r^2}{Q_1Q_5}dy^2\,,\label{eq:AdSmetric}\\
ds^2_{S^3}&=d \theta^2+\sin^2\theta\, d\hat{\phi}^2+\cos^2\theta\, d\hat{\psi}^2\,.
\end{align}
\end{subequations}
The coordinates $t, y$ are identified with the time and space coordinates of the CFT, and we take $y$ to parametrize an $S^1$ of radius $R$; $\hat \phi$ and $\hat \psi$ are some linear combinations of the $S^3$ Cartan's angles $\phi$, $\psi$ and the CFT coordinates $t$, $y$; the particular linear combination depends on the state and will be given below. $Q_1$ and $Q_5$ encode the numbers of D1 and D5 charges, $n_1$ and $n_5$ (with $N=n_1 n_5$):
\begin{equation}
Q_1 = \frac{(2\pi)^4 n_1 g_s (\alpha')^3}{V_4}\,,\quad Q_5 =  g_s n_5 \alpha'\,,
\end{equation}
where $g_s$ is the string coupling and $V_4$ is the volume of the compact space $M$. The parameter $a$ is linked to the D-brane charges and the $S^1$ radius by
\begin{equation}
a=\frac{\sqrt{Q_1 Q_5}}{R}\,.
\end{equation}
Finally $k$ is a positive integer which introduces a conical defect in the geometry $ds^2_{AdS_3}$: indeed this space represents a $\mathbb{Z}_k$ orbifold of AdS$_3$.

The gravity solution also includes a RR 2-form, whose field strength is
\begin{subequations}\label{eq:F3}
\begin{align}
F_3&=2\, Q_5\,(-\mathrm{vol}_{AdS_3}+\mathrm{vol}_{S^3})\,,\\
\mathrm{vol}_{AdS_3}&=\frac{r}{Q_1Q_5}\,d r\wedge dt \wedge dy\,,\quad \mathrm{vol}_{S^3}=\sin\theta \cos\theta\, d \theta\wedge  d\hat{\phi} \wedge d\hat{\psi}\,.
\end{align}
\end{subequations}
The 3-form field strength is anti-self-dual in the 6D Einstein metric
\begin{equation}
*_6 F_3 = - F_3\,,
\end{equation}
where $*_6$ is the Hodge star with respect to $ds^2$ and we choose the orientation $\epsilon_{r t y \theta \hat \phi \hat \psi}=+1$.

\subsubsection{The two-charge states}
 
The states $|s=0,k\rangle$ have $h_H=\bar h_H = \frac{c}{24}=\frac{N}{4}$ and thus carry D1 and D5 charges but no momentum charge. The geometries dual to these states were found in \cite{Lunin:2001jy} and can be written in the form \eqref{eq:6Dmetric} with 
\begin{equation}\label{eq:SF2charge}
\hat \phi = \phi - \frac{t}{R\,k}\,,\quad \hat \psi = \psi - \frac{y}{R\,k}\,.
\end{equation}
Note that the original set of coordinates $(t,y,\phi,\psi)$ is subject to the identifications 
\begin{equation}
(t,y,\phi,\psi)\sim (t,\,y+ 2\pi\,l\, R ,\, \phi+2\pi\,m ,\, \psi + 2\pi\,n)\,,
\end{equation}
with $l , m, n \in \mathbb{Z}$. Only when $k=1$ eq. \eqref{eq:SF2charge} defines a new set of coordinates $(t,y,\hat\phi,\hat\psi)$ which satisfy analogous identifications  
\begin{equation}\label{eq:newidentifications}
(t,y,\hat\phi,\hat\psi)\sim (t,\,y+ 2\pi\,l\, R ,\, \hat\phi+2\pi\,m ,\, \hat\psi + 2\pi\,n)\,,
\quad (k=1)\,.
\end{equation}
In this case the coordinate transformation $(t,y,\phi,\psi)\to (t,y,\hat\phi,\hat\psi)$ realizes the spectral flow from the state  $|s=0,k=1\rangle$ to the SL(2,$\mathbb{C}$)-invariant vacuum, whose dual geometry is \eqref{eq:6Dmetric} with the identifications \eqref{eq:newidentifications}, i.e. global AdS$_3\times S^3$.  For $k>1$ the identifications induced on the $(t,y,\hat\phi,\hat\psi)$ coordinates are more complicated:
\begin{equation}
(t,y,\hat\phi,\hat\psi)\sim \left (t,\,y+ 2\pi\,l\, R ,\, \hat\phi+2\pi\,m ,\, \hat\psi -2\pi\,\frac{l}{k}+ 2\pi\,n\right )\,.
\end{equation}
The geometry dual to the state $|s=0,k\rangle$ is given by \eqref{eq:6Dmetric} expressed in the $(t,y,\phi,\psi)$ coordinate system via \eqref{eq:SF2charge}: geometrically it represents a $\mathbb{Z}_k$ orbifold of  AdS$_3\times S^3$. For $k>1$ there is no state in the D1D5 CFT dual to the geometry \eqref{eq:6Dmetric} with the identifications \eqref{eq:newidentifications}.

\subsubsection{The three-charge states}
The states $|s,k\rangle$ have $h_H=\frac{N}{4} + \frac{N\,s(s+1)}{k^2}$, $\bar h_H=\frac{N}{4}$ and thus carry momentum $n_p = h-\bar h = \frac{N\,s(s+1)}{k^2}$. The dual geometries have been found in \cite{Giusto:2012yz} and are of the form \eqref{eq:6Dmetric} with 
\begin{equation}\label{eq:SF3charge}
\hat \phi = \phi - \frac{t}{R\,k}-s\,\frac{t+y}{R\,k}\,,\quad \hat \psi = \psi - \frac{y}{R\,k}-s\,\frac{t+y}{R\,k}\quad (s\in \mathbb{Z})\,.
\end{equation}
As in the previous example, this coordinate redefinition preserves the simple periodic identifications only for $k=1$. For $k>1$ the geometry is again a $\mathbb{Z}_k$ orbifold of  AdS$_3\times S^3$, though the orbifold group, determined by the coordinate redefinition \eqref{eq:SF3charge}, acts differently than in the previous example. It is important to keep in mind that the integers $s$ and $k$ must be such that the momentum on each strand $s(s+1)/k$ be integer\footnote{This condition only holds when $n_1$ and $n_5$ are coprime and a more general condition applies if $n_1$ and $n_5$ share a common divisor~\cite{Giusto:2012yz}; for simplicity we will assume $n_1$ and $n_5$ coprime  in this article.}. This allows for non-integer $s/k$; states with $s/k$ integer are particularly simple,  as they are obtained from the 2-charge states with $s=0$ by a global chiral algebra transformation.

We note that setting $s=0$ the D1D5P states specified by eq. \eqref{eq:SF3charge} reduce to the D1D5 states corresponding to \eqref{eq:SF2charge}. In the following we will thus work with the more general class of states described by \eqref{eq:SF3charge}.

\subsection{The holographic two-point function}
We want to compute the correlator of the light operators $O_L\equiv O^{++}$ and ${\bar O}_L\equiv O^{--}$ in the states $|s,k\rangle$, whose dual geometries are specified by (\ref{eq:6Dmetric},\ref{eq:F3}) and \eqref{eq:SF3charge}. We will do this by computing the vev of the operator ${\bar O}_L$ in the presence of a source for the operator $O_L$, and then differentiating the vev with respect to the source to obtain the two-point correlator:
\begin{equation}\label{eq:twopoint}
\langle s,k | O_L(0,0) {\bar O}_L(t,y) |s,k \rangle =\ii \frac{\delta \langle {\bar O}_L(t,y)  \rangle_J}{\delta {\bar J}_L(0,0)} \Bigl |_{J=0}\,, 
\end{equation}
where ${\bar J}_L$ is the source coupling to $O_L$ and the correlator is computed on the cylinder parametrized by $t$ and $y$. The vev $\langle {\bar O}_L(t,y)  \rangle_J$ is extracted from the supergravity field dual to ${\bar O}_L$. 

In 6D\footnote{When lifted to the 10D IIB duality frame, $B_2$ is the NSNS 2-form and $w$ is the component of the RR 4-form along the compact space $M$.} the fields dual to the chiral primary operators $O^{\pm \pm}$ are a scalar $w$ and a 2-form $B_2$, which satisfy a coupled system of differential equations. The linearization of these equations around the background (\ref{eq:6Dmetric},\ref{eq:F3}) gives \cite{Deger:1998nm,Mathur:2003hj}
\begin{equation}\label{eq:6Deqs}
d B_2 - *_6 d B_2 = 2\,w\,F_3\,,\quad d *_6 d w = \frac{Q_1}{Q_5}\,dB_2\wedge F_3\,.
\end{equation}
The factorised form of the background (when expressed in $\hat \phi$, $\hat \psi$ coordinates) allows to reduce the 6D equations \eqref{eq:6Deqs} to two sets of decoupled equations on AdS$_3$ and $S^3$. To this purpose one can make the ansatz~\cite{Shigemori:2013lta}
\begin{equation}
w = Y\,B\,,\quad B_2 = \gamma\, (Y\,*_{AdS_3} \! dB - B\,*_{S^3} \! dY)\,,
\end{equation}
where $Y$ is a function of $\theta, \hat \phi, \hat \psi$, $B$ is a function of $r, t, y$, $*_{AdS_3}$ and $*_{S^3}$ are the Hodge duals with respect to $ds^2_{AdS_3}$ and $ds^2_{S^3}$ and $\gamma$ is a constant that will be determined shortly. It is straightforward to verify that this ansatz satisfies \eqref{eq:6Deqs} if $Y$ and $B$ are eigenfunctions of the respective Laplacians:
\begin{equation}\label{eq:3Deqs}
\Box_{AdS_3} B = \ell (\ell-2)\,B \,,\quad \Box_{S^3} Y = -\ell(\ell+2)\,Y\,,
\end{equation}
and if $\gamma=\frac{Q_5}{\ell}$.
Then $Y$ is a scalar harmonic on $S^3$ of order $\ell$, with $\ell$ a positive integer; $B$ is a minimally coupled scalar in AdS$_3$ with mass $m^2=\ell(\ell-2)$.

As the CPO's $O^{\pm\pm}$ form a multiplet with $SU(2)_L \times SU(2)_R$ charges $j= \bar j=1/2$, the gravity dual field must have spin 1, and hence we should look for solutions for $B$ and $Y$ with $\ell=1$. The vev of $O^{--}$ is encoded in the component of the field $w$ proportional to the spherical harmonic $Y_1^{++}= \sin\theta \, e^{\ii \phi}$ (see eqs. (4.9), (4.10) in \cite{Giusto:2015dfa}). Thus we seek for a solution of the form
\begin{equation}
w = B(t,y,r)\, \sin\theta \, e^{\ii \hat\phi} = B(t,y,r)\, e^{-\ii\, \frac{t}{R k} - \ii \,s\,\frac{t+y}{R k}}\,\sin\theta \, e^{\ii\phi} \,,
\end{equation}
where $B(t,y,r)$ solves the AdS$_3$ Laplace equation \eqref{eq:3Deqs} with $\ell=1$. Note that the phase $e^{- \ii \,s\,\frac{y}{R k}}$ is not globally well-defined on the circle $y\sim y + 2\pi\,R$ when $s/k$ is fractional. Thus, for $w$ to be a globally defined field, we need to require that the function $B(t,y,r)$ has an appropriate monodromy when going around the $S^1$ to cancel that of the phase:
\begin{equation}
\label{eq:monodromy}
B(y,y+2\pi R,r) = B(t,y,r)\,e^{\ii\, \frac{\hat s}{k} \,2\pi}\,,
\end{equation}
where $\hat s = s\, \mathrm{mod} \,k$ and we choose $0\le \hat s <k$.

Since the non-normalizable and normalizable solutions of the AdS$_3$ wave equation go like $r^{-1} \log r$ and $r^{-1}$, the usual AdS/CFT prescription implies that the asymptotic behaviour of the field $w$ has the form
\begin{equation}
w \approx \frac{{\bar J}_L(t,y)\,\log r + \langle {\bar O}_L(t,y)  \rangle_J }{r}\sin\theta \, e^{\ii\phi} \,.
\end{equation}
Requiring that $w$ is finite in the interior of space links the normalizable and non-normaliz\-able terms of the solution. In accordance with \eqref{eq:twopoint}, the two point function of $O_L(0,0)$ and $\bar O_L(t,y)$ is given by the vev  $\langle {\bar O}_L(t,y)  \rangle_J$ when the source for $O_L$ is a delta-function: ${\bar J}_L(t,y)= \delta(t,y)$.

In summary, one looks for a solution of the equation \eqref{eq:3Deqs} for $B$ with $\ell=1$ which is {\it regular} in the bulk, has the monodromy \eqref{eq:monodromy}, and its leading behavior at large $r$ is
\begin{equation}
\label{eq:sol1}
B(t,y,r)\approx \delta(t,y) \,\frac{\log r}{r} + b_1(t,y)\,\frac{1}{r}\,.
\end{equation}
AdS solutions with monodromies like in~\eqref{eq:monodromy} are not usually considered in the literature. In Appendix \ref{sec:waveequation} we will derive the solution of the wave equation in AdS$_3/\mathbb{Z}_k$ with the boundary conditions prescribed above by generalising the computations in \cite{Skenderis:2008dg,Aref'eva:2016pew}. One finds
\begin{equation}
\label{eq:scalar}
b_1(t,y) =-\ii \frac{e^{\ii \hat s \frac{y}{R\,k}}}{e^{\ii \frac{t}{R\,k}} - e^{-\ii \frac{t}{R\,k}}}\,\left[\frac{e^{\ii \frac{t-y}{R}}}{e^{\ii \frac{t-y}{R}}-1}e^{-\ii \hat s\frac{t}{R\,k}}+\frac{1}{e^{\ii \frac{t+y}{R}}-1} e^{\ii \hat s \,\frac{t}{R\,k}}\right]\,.
\end{equation}
The two-point correlator of the light operators in the state $|s,k\rangle$ is given by
\begin{equation}\label{eq:bulkcorrelator}
\langle  s,k | O_L(0,0) {\bar O}_L(t,y) |s,k \rangle =  \ii \,b_1(t,y)\, e^{-\ii \,\frac{t}{R k} - \ii \,s\,\frac{t+y}{R k}}\,.
\end{equation}

To compare the bulk result \eqref{eq:bulkcorrelator} with the CFT, one should transform from the cylinder coordinates $t$ and $y$ to the Euclidean plane coordinates\footnote{This is different from what is done when the thermal results are extracted from the Euclidean correlators. Of course in the thermal case, one needs to perform the Wick rotation so as to identify the compact coordinate with time and, on the bulk side, the four point correlators are compared with the wave equation on a BTZ black hole.} $z, \bar z$:
\begin{equation}
z= e^{\ii\,\frac{t+y}{R}}\,,\quad \bar z = e^{\ii\,\frac{t-y}{R}}\,,
\end{equation}
and remember that 
\begin{equation}
O_L(z,\bar z)= (z \bar z)^{-1/2}\,O_L(t,y)\,,
\end{equation}
(and the same for ${\bar O}_L$) since $O_L(z,\bar z)$ is a primary of dimension $h_L=\bar h_L =1/2$. The gravity result for the correlator on the plane is then
\begin{equation}\label{eq:bulkcorrelatorplane}
\langle  s,k | O_L(1) {\bar O}_L(z,\bar z) |s,k \rangle =\frac{z^{\frac{\hat s-s}{k}}}{|z| \,|1-z|^2}\frac{1-|z|^{2(1-\frac{\hat s}{k})} + \bar z \,(|z|^{-2\frac{\hat s}{k}}-1)}{1-|z|^{\frac{2}{k}}}\,.
\end{equation}
One can check that when $s=k p$ (and thus $\hat s=0$) the previous result reduces to the CFT expression \eqref{eq:resultsfk}, and when $s=k p -1$ (and thus $\hat s=k-1$) one recovers  \eqref{eq:resultsfk-1}, up to overall numerical coefficients that have not been kept in the gravity derivation.

%%%%%%%%%%%%%%%%%%%%%%%%%%%%%%%%%%%%%%%%%

\section{Discussion}
\label{sec:discussion}

It is well known that symmetric orbifolds provide a prototypical example of CFTs that have a sparse spectrum, which is a necessary condition to have a dual gravitational description in terms of a string or supergravity theory~\cite{Heemskerk:2009pn}. We focused on the best known example of such orbifold theories, the D1D5 CFT at the free point. In section~\ref{sec:cfts} we calculated on the CFT side a very special class of $4$-point correlators among BPS operators, where two states are heavy (i.e. have conformal dimension of order $c$), while the other two are light (i.e. their conformal dimension is of order $1$).  These correlators are essentially combination of the free-fermion result and, in the $(O_H O_H) (O_L O_L)$ OPE, are completed saturated by the affine identity block of a $U(1)$ subgroup of the $SU(2)$ symmetry of the theory. This suggests that they are protected by supersymmetry and motivates the supergravity analysis of section~\ref{sec:grav}. Again thanks to the simplicity of our external states, also the gravity calculation is easy and, in this case, the basic ingredient is obtained by studying the scalar wave equation in AdS$_3/\mathbb{Z}_k$. Then in order to obtain the full correlator it is important to know how the 3D result is uplifted to the full 10D geometry. In all examples under analysis, we find agreement with the free CFT result, even if this description is valid in different point of the moduli space, thus confirming the expectations based on supersymmetry as mentioned above.

Of course, in the Euclidean case, the correlators we studied are singular only in the OPE limits. One of the main features of our result is that, for the whole correlator, this holds even at the leading order in the large $c$ limit, while, in the same limit, the contribution of the Virasoro identity block in the $(O_H O_H) (O_L O_L)$ OPE develops spurious singularities~\cite{Fitzpatrick:2015dlt,Fitzpatrick:2016ive}. In other words, the $c\to\infty$ limit of the correlators studied here is not captured by the contribution of the identity Virasoro block in the heavy-light channel. This is reflected by the gravity calculations: the 2-point functions of the light operators in the near-horizon limit of the Strominger-Vafa black hole (which is the extremal BTZ) captures just the identity Virasoro block, while the same calculation in the microstate geometry dual to the heavy state reproduces the whole 4-point correlators, including the contributions of the higher order Virasoro primaries. This supports the intuition that the black hole geometry describes the correlators in a statistical ensemble, while each individual microstate yields correlators that deviate from the statistical answer before one reaches singularities that are usually related to the presence of a horizon.

In our case, due to the simple form of the heavy states, these deviations are present even at distances larger than the Schwarzschild radius. On the CFT side, this means that, in the $(HH)(LL)$ OPE, there are contributions of non-trivial Virasoro primaries with conformal dimension of order $1$. The pattern discussed above is different from the one advocated in~\cite{Fitzpatrick:2016ive}, where it is suggested that quantum (i.e. $1/c$ corrections) are needed to resolve the spurious singularities of the statistical/black-hole result. Thus it is natural to ask whether the regularity of our Euclidean correlators in the large $c$ regime is due to some peculiar feature of the D1D5 CFT under analysis and/or is a consequence of the very special operators considered. We believe that this is actually a general property as argued below.

The absence of spurious singularities at finite values of the central charge $c$ is a direct consequence of the convergence of the OPE expansion in unitary CFT and of the basic properties of the Hilbert space structure of the spectrum~\cite{Pappadopulo:2012jk}. In a nutshell, in the radial quantization, one can separate the four operators in the correlator by a sphere of radius $r$, with $|z_4|<|z_3|<r<|z_2|<|z_1|$. Then the convergence of the OPE ensures that the operators $O_1$ and $O_2$ in the external region produce a new state $|\phi_e\rangle$ on the sphere and the same happens, in the internal region, for the operators $O_3$ and $O_4$ that produce $|\phi_i\rangle$ (of course if $z_1\to\infty$, $z_2=1>z_3>z_4=0$, $|\phi_i\rangle$ depends on $z=1-z_3$). So the $4$-point correlator reduces to the scalar product $\langle \phi_e |\phi_i(z) \rangle$ which is finite for any value of $z$ in the interval $0<|z|<1$. In~\cite{Fitzpatrick:2016ive} it was noted that it is not straightforward to take the $c\to\infty$ limit in this argument if one identifies $O_1$, $O_2$ with the heavy operators and $O_3$, $O_4$ with the light ones. We can see this directly in the simplest one of our examples, i.e. the correlator with the operators~\eqref{eq:Opp} and~\eqref{eq:OLM}. The OPE between the light operators reads
\begin{equation}
  \label{eq:opel}
  O_L(w) \bar{O}_L(0) = \frac{1}{|w|^2} + \frac{1}{N} \sum_r \left( \frac{J^3_{(r)}}{\bar{w}} + \frac{\tilde{J}^3_{(r)}}{w} \right) + \frac{1}{N} \sum_{r\not= s} O^L_{(r)} O^L_{(s)} + \ldots
\end{equation}
In the large $c$ limit, normally one would discard the contribution of the terms with the currents, as their norm is of order $1/N$. However the OPE between the heavy operators produces terms, again proportional to the currents, that are non-normalizable in the $N\to \infty$ limit
\begin{equation}
  \label{eq:opeh}
  O_H(w) \bar{O}_H(0) =  \frac{1}{|w|^{2 h_H}} \left(1 + w \sum_r  J^3_{(r)} +  \bar{w} \sum_r  \tilde{J}^3_{(r)} + \ldots  \right)\,.
\end{equation}
Such non-normalizable terms can combine with the currents that appear in \eqref{eq:opel} to give non-negligible contributions to the block decomposition of the correlator; moreover their presence invalidates the regularity argument based on the existence of a well-defined scalar product, and is probably responsible for the singular behaviour of the heavy-light Virasoro blocks.

At the level of the correlators one can repeat the same derivation focusing on the OPE channel where the light operators are close to the heavy ones. In this case the intermediate states are normalizable even in the $c\to \infty$ limit and so the argument discussed above shows that the large $c$ Euclidean correlators should not have spurious singularities. Of course this does not provide any information on the identity Virasoro block nor other $(HH)(LL)$ blocks because they do not appear in the $(HL)(HL)$ decomposition. However once the regularity of the large $c$ limit of the correlators is established, we know that there is an infinite number of Virasoro primaries contributing to the $(HH)(LL)$ OPE. In the simple cases considered in this paper, it turns out that these primaries are protected, as they are affine descendants of the identity operator. Thus the correlator we compute at the CFT orbifold point reproduces the one extracted from the dual geometry: in these instances then correlators are regular already at the level of supergravity. In general the OPE argument in the $(HL)(HL)$ channel predicts that correlators be regular in the large $c$ limit {\it at a generic point} in the CFT moduli space. We do not expect, however, that all the operators ensuring the absence of spurious singularities at large $c$ will be captured in the supergravity approximation. It would be an important progress to identify explicitly the CFT operators that are relevant to the $(HH)(LL)$ decomposition of a more general correlator. This could help to understand from a CFT prospective what contributions survive in the large $c$ limit beside those that reproduce the thermal behaviour. 

It is of course very interesting to elucidate the meaning of this pattern on the dual gravity side, where the main question is whether there are effects that modify the standard general relativity picture at the scales of Schwarzschild radius $R_s$ in the limit where $R_s$ is large in Planck units. Scenarios that fall in this class are the fuzzball~\cite{Mathur:2005zp,Mathur:2008nj} and the firewall~\cite{Almheiri:2012rt} proposals. In situations that can be studied within the AdS/CFT duality, one could rephrase these ideas by saying that the heavy-light correlators, in a pure heavy state should differ from the ones calculated in a statistical ensemble even in the $c\to\infty$ limit. This is exactly the behaviour we observe in the simple correlators analyzed in this paper. Of course, even if this is a general pattern as suggested above, there are several points that need to be understood in order to have a complete picture on the gravitational side. These include the following questions: what are the non-trivial operators that generically appear in $(HH)(LL)$ decomposition of a {\em typical} heavy states? Is it possible to associate a scale in the radial direction to these contributions and show it is of the same order of $R_s$? For which correlators the contributions from non-trivial conformal blocks are negligible and the result is well approximated by the thermal correlator? Posing such questions in this framework might help to clarify some aspects of the ``fuzzball complementarity'' conjecture~\cite{Mathur:2011wg,Mathur:2012jk}.

We conclude by discussing some less speculative and more concrete possible developments. Of course it would be interesting to consider $4$-point correlators that are not related by a change of coordinates to $2-$point functions in AdS$_3/\mathbb{Z}_k$. In the same spirit, also changing the form of the light operators could provide new information on how different objects probe the heavy backgroud. Both these generalizations would allow to compare the bulk and the CFT results in examples with a richer structure. Another interesting direction is to exploit the relation between the D1D5 CFT and gravitational theories with a higher spin symmetry, as discussed in~\cite{Gaberdiel:2014cha,Gaberdiel:2015uca,Gaberdiel:2015wpo}. It seems possible to study heavy-light correlators where the operators belong to the pure gravitational Vasiliev sector and it would be interesting to compare this bulk description with the CFT result at the free orbifold point (something that could now be done without relying on non-renormalization theorems). Finally it would be interesting to analyze heavy-light $4$-point correlators in other CFTs that have a holographic interpretation at large $c$, so as to check or disprove the generality of the pattern suggested by the analysis for the D1D5 CFT.

\vspace{7mm}
 \noindent {\large \textbf{Acknowledgements} }

 \vspace{5mm} 

We would like to thank M. Beccaria, K. Skenderis, A. Veliz-Osorio for useful discussions and correspondence. This research is partially supported by STFC (Grant ST/L000415/1, {\it String theory, gauge theory \& duality}), by the Padova University Project CPDA119349 and by INFN.

%%%%%%%%%%%%%%%%%%%%%%%%%%%%%%%%%%%%%%%%%

\appendix

\section{The CFT at the orbifold point}
\label{sec:cftconv}
In this appendix we will state what the ingredients for the CFT computations are and how they are put together to get the results in section \eqref{sec:cfts}. Let's start from the untwisted sector of the theory: the holomorphic (left) fermions on the $r$-th strand $\psi_{(r)}^{\alpha\dot{A}}$ and the antiholomorphic (right) ones $\tilde{\psi}_{(r)}^{\dot{\alpha}\dot{A}}$ have the nontrivial OPEs
\begin{align}
 \label{eq:fermionsope}
\psi_{(r)}^{+\dot{A}}(z) \psi_{(r)}^{-\dot{B}}(w) = \frac{\epsilon^{\dot{A}\dot{B}}}{z-w} + \left[\mathrm{Reg.}\right],\qquad \tilde{\psi}_{(r)}^{+\dot{A}}(\bar{z}) \tilde{\psi}_{(r)}^{-\dot{B}}(\bar{w}) = \frac{\epsilon^{\dot{A}\dot{B}}}{\bar{z}-\bar{w}} + \left[\mathrm{Reg.}\right],
\end{align}
where our convention is $\epsilon_{\dot{1}\dot{2}} = - \epsilon^{\dot{1}\dot{2}}=1$. The indices $\alpha, \dot{\alpha}$ are in the fundamental representation of $SU(2)$ and will take values $\alpha, \dot{\alpha}=\pm$ or $\alpha, \dot{\alpha}=1,2$ depending on what's more convenient case by case. Through bosonization, the fermions can be written in terms of bosons $H(z), K(z)$ as
\begin{equation}
 \label{eq:bosonization}
\psi_{(r)}^{+\dot{1}} = \ii\ e^{\ii H_{(r)}},\quad \psi_{(r)}^{-\dot{2}} = \ii\ e^{-\ii H_{(r)}},\quad \psi_{(r)}^{+\dot{2}} = e^{\ii K_{(r)}},\quad \psi_{(r)}^{-\dot{1}} = e^{-\ii K_{(r)}},
\end{equation}
and an analogous dictionary holds for the right fermions, with bosons $\tilde{H}_{(r)}(\bar{z}), \tilde{K}_{(r)}(\bar{z})$. The bosons have the nontrivial OPEs
\begin{equation}
 \label{eq:bosonsope}
H_{(r)}(z) H_{(r)}(w) = -\log\left(z-w\right) + \left[\mathrm{Reg.}\right],\qquad K_{(r)}(z) K_{(r)}(w) = -\log\left(z-w\right) + \left[\mathrm{Reg.}\right],
\end{equation}
and the rule for contractions of bosonized fields is
\begin{equation}
 \label{eq:boscontr}
:e^{\ii \alpha H_{(r)}(z)}: :e^{\ii \beta H_{(r)}(w)}: = \left( z-w \right)^{-\alpha\beta}\ :\exp\bigl( \alpha H_{(r)}(z) + \beta H_{(r)}(w) \bigr):.
\end{equation}

A further ingredient we need is given by the current operators. In the untwisted sector, these are written as
\begin{equation}
 \label{eq:Jtot}
J^a(z) = \sum_{r=1}^N J^a_{(r)}(z)\,,
\end{equation}
with
\begin{subequations}
 \label{eq:Jr}
\begin{align}
J^3_{(r)} &= -\frac{1}{2} \left(\psi_{(r)}^{+\dot{A}} \psi_{(r)}^{-\dot{B}} \epsilon_{\dot{A}\dot{B}}\right),\\
J^+_{(r)} &= J^1_{(r)} + \ii J^2_{(r)} = \frac{1}{2} \psi_{(r)}^{+\dot{A}} \psi_{(r)}^{+\dot{B}}\epsilon_{\dot{A}\dot{B}}\,,\\
J^-_{(r)} &= J^1_{(r)} - \ii J^2_{(r)} = - \frac{1}{2} \psi_{(r)}^{-\dot{A}} \psi_{(r)}^{-\dot{B}}\epsilon_{\dot{A}\dot{B}}\,.
\end{align}
\end{subequations}
Since the theory enjoys an $SU(2)_1\times U(1)$ affine symmetry on each strand, we also have a $U(1)$ generator $J^0$ defined as
\begin{equation}
J^0(z) = \sum_{r=1}^N J^0_{(r)}(z)\,,\qquad J^0_{(r)} = -\frac{C}{2} \psi_{(r)}^{\alpha\dot{1}}\psi_{(r)}^{\beta\dot{2}}\epsilon_{\alpha\beta}\,,
\end{equation}
where $C$ is a constant that is not fixed by the algebra (corresponding to the fact that the level of the $U(1)$ factor inside a $SU(2)_k\times U(1)$ affine algebra is undetermined). $J^3_{(r)}$ can also be written in terms of the bosons $H$ and $K$, noticing that
\begin{equation}
 \label{eq:psi2bos}
\psi_{(r)}^{+\dot{1}} \psi_{(r)}^{-\dot{2}} = -\ii \partial H_{(r)},\qquad \psi_{(r)}^{+\dot{2}} \psi_{(r)}^{-\dot{1}} = \ii \partial K_{(r)},
\end{equation}
as
\begin{equation}
 \label{eq:Jbos}
J^3_{(r)} = \frac{\ii}{2} \bigl( \partial H_{(r)} + \partial K_{(r)} \bigr),\qquad J^+_{(r)} = \ii e^{\ii \left( H_{(r)} + K_{(r)}\right)},\qquad J^-_{(r)} = \ii e^{-\ii \left( H_{(r)} + K_{(r)}\right)}.
\end{equation}

The light operators we consider are (on a strand)
\begin{equation}
 \label{eq:lightopuntwist}
O^L_{(r)} = -\frac{\ii}{\sqrt{2}} \psi^{+\dot{A}}_{(r)} \epsilon_{\dot{A} \dot{B}} \tilde{\psi}^{+ \dot{B}}_{(r)} \equiv O^{++}_{(r)}~,\qquad \bar{O}^L_{(r)} = \frac{\ii}{\sqrt{2}} \psi^{-\dot{A}}_{(r)} \epsilon_{\dot{A} \dot{B}} \tilde{\psi}^{- \dot{B}}_{(r)} \equiv O^{--}_{(r)}~,
\end{equation}
while the ones acting on the product theory are given by the sum over copies in \eqref{eq:HLO}. In all the cases considered we will get the same result for each strand, so we can just work on a copy and (in the untwisted sector) multiply by $N$. The heavy operators that we want to consider for the correlators in the untwisted sector are obtained from~\eqref{eq:HLO} and~\eqref{eq:OLMs} with
\begin{equation}
  \label{eq:Spines}
  S^{\dot{1}}_{s, (r)} = {\rm e}^{\ii (s + \frac{1}{2}) H_{(r)}}~,~~
  S^{\dot{2}}_{s, (r)} = {\rm e}^{\ii (s + \frac{1}{2}) K_{(r)}}~.
\end{equation}
The corresponding states are
\begin{equation}
\begin{aligned}
|s,k=1\rangle &\equiv \lim_{z,\bar z\to 0} O^H(s,k=1;z,\bar z)|0\rangle \\
&=\otimes_{r} \left[(J^+_{-2s})_{(r)}\ldots (J^+_{-2})_{(r)} \lim_{z,\bar z\to 0} O^H_{(r)}(s=0,k=1;z,\bar z)\right]|0\rangle \,.
\end{aligned}
\end{equation}
The left and right parts of the four-point function \eqref{eq:HLO} factorize and we need to evaluate corelators of the form
\begin{align}
 \label{eq:corrcompuntwist}
F_{s,(r)}^{\dot{A}\dot{C}}(z_i) &\equiv \langle e^{\ii \left( s +\frac{1}{2}\right) \left( H_{(r)}(z_1) + K_{(r)}(z_1)\right)} e^{-\ii \left( s +\frac{1}{2}\right) \left( H_{(r)}(z_2) + K(z_2)_{(r)}\right)} \psi_{(r)}^{+\dot{A}}(z_3) \psi_{(r)}^{-\dot{C}}(z_4)\rangle\times\nonumber\\
&\quad\times\prod_{r^\prime \neq r} \langle e^{-\ii \left(s + \frac{1}{2}\right)\left(H_{(r)}(z_1) + K_{(r)}(z_1)\right)} e^{\ii \left(s + \frac{1}{2}\right)\left(H_{(r)}(z_2) + K_{(r)}(z_2)\right)}\rangle.
\end{align}
The right part is completely analogous, with the exception that in the right sector we always have $s=0$. Notice that in principle the light operators acting on the product theory bring two sums over strands. Despite this, by spin conservation, the only nonzero contributions come from the cases in which both light operators act on the same strand, which reduces the full correlator to just one sum over copies. Moreover, since the heavy operators are product over copies, the term relative to the $r$-th copy is multiplied by the two-point functions of the heavy operators on all the copies $r^\prime \neq r$. The full correlation function reads
\begin{equation}
 \label{eq:corruntwist}
\langle O_H(z_1) \bar{O}_H(z_2) O_L(z_3) \bar{O}_L(z_4) \rangle = \sum_{r=1}^N  \frac{1}{2}\ F_{s,(r)}^{\dot{A}\dot{C}}(z_i) F_{0,(r)}^{\dot{B}\dot{D}}(\bar{z}_i)\ \epsilon_{\dot{A}\dot{B}} \epsilon_{\dot{C}\dot{D}} .
\end{equation}
$F_{s,(r)}^{\dot{A}\dot{C}}(z_i)$ is nonzero only if the two fermions can have a nontrivial contraction, which selects the cases $(\dot{A}, \dot{B}) = (\dot{1}, \dot{2})$ and $(\dot{A}, \dot{B}) = (\dot{2}, \dot{1})$. In the first case, using \eqref{eq:boscontr} to contract each possible pair of fields, we get
\begin{align}
 \label{eq:F12untwist}
F_{s,(r)}^{\dot{1}\dot{2}}(z_i) = - \frac{z_{13}^{s+\frac 12} z_{24}^{s+\frac 12}}{z_{12}^{2 h} z_{14}^{s + \frac 12} z_{23}^{s + \frac 12} z_{34}}=-\frac{1}{z_{12}^{2h} z_{34}} \,z^{-s-\frac{1}{2}}\,,
\end{align}
where $h = \left(s+\frac{1}{2}\right)^2$. The second case is analogous, giving $F_{s,(r)}^{\dot{2}\dot{1}}(z_i) = - F_{s,(r)}^{\dot{1}\dot{2}}(z_i)$. The antiholomorphic parts are obtained from these setting $s=0$ and replacing $z_i\to \bar{z}_i$ and $h\to\bar{h} = 1/4$. Putting everything together we get
\begin{equation}
 \label{eq:resultuntwist}
\langle O_H(z_1) \bar{O}_H(z_2) O_L(z_3) \bar{O}_L(z_4) \rangle = \frac{1}{z_{12}^{2h_H} \bar{z}_{12}^{2\bar{h}_H} |z_{34}|^2} \,|z|^{-1}\,z^{-s}\,;
\end{equation}
a factor $N$ would come from the fact that each term of the sum over $r$ gives the same contribution, but this is cancelled by the normalization \eqref{eq:HLO} of $O_L$. The first correlator we compute in the untwisted sector corresponds to $s=0$, while the second to generic $s$.

Let's consider the twisted sector. In this case we have $N/k$ strands of length $k$, each of which contains $k$ left fermions $\psi_{(r)}^{\alpha\dot{A}}$ and $k$ right fermions $\tilde{\psi}_{(r)}^{\dot{\alpha}\dot{A}}$, with $r=1,\ldots,k$. If we rotate $z\to e^{2\pi \ii}\ z$ we have that
\begin{equation}
 \label{eq:fermionlengthk} 
\psi_{(r)}^{\alpha\dot{A}}(e^{2\pi \ii}\ z) = \psi_{(r+1)}^{\alpha\dot{A}}(z)\quad\forall r=1,\ldots,k-1,\qquad \psi_{(k)}^{\alpha\dot{A}}(e^{2\pi \ii}\ z) = \psi_{(1)}^{\alpha\dot{A}}(z).
\end{equation}
We can diagonalize this periodicity condition by defining new fermions
\begin{equation}
 \label{eq:fermionsrho}
\begin{aligned}
\psi_{\rho}^{+\dot{A}}(z) = \frac{1}{\sqrt{k}} \sum_{r=1}^k e^{2\pi\ii \frac{r\rho}{k}}\ \psi_{(r)}^{+\dot{A}}(z),\qquad \tilde{\psi}_{\rho}^{\dot{+}\dot{A}}(\bar{z}) = \frac{1}{\sqrt{k}} \sum_{r=1}^k e^{-2\pi\ii \frac{r\rho}{k}}\ \tilde{\psi}_{(r)}^{\dot{+}\dot{A}}(\bar{z}),\\
\psi_{\rho}^{-\dot{A}}(z) = \frac{1}{\sqrt{k}} \sum_{r=1}^k e^{-2\pi\ii \frac{r\rho}{k}}\ \psi_{(r)}^{-\dot{A}}(z),\qquad \tilde{\psi}_{\rho}^{\dot{-}\dot{A}}(\bar{z}) = \frac{1}{\sqrt{k}} \sum_{r=1}^k e^{2\pi\ii \frac{r\rho}{k}}\ \tilde{\psi}_{(r)}^{\dot{-}\dot{A}}(\bar{z}),
\end{aligned}
\end{equation}
which behave like
\begin{equation}
 \label{eq:diagonalizedperiod}
\psi_{\rho}^{+\dot{A}}(e^{2\pi\ii}\ z) = e^{-2\pi\ii \frac{\rho}{k}}\ \psi_{\rho}^{+\dot{A}}(z),\qquad \tilde{\psi}_{\rho}^{\dot{+}\dot{A}}(e^{-2\pi\ii}\ \bar{z}) = e^{2\pi\ii \frac{\rho}{k}}\ \tilde{\psi}_{\rho}^{\dot{+}\dot{A}}(\bar{z}),
\end{equation}
where the behaviour of $\psi_{\rho}^{-\dot{2}}$ and $\psi_{\rho}^{-\dot{1}}$ are obtained respectively from $\psi_{\rho}^{+\dot{1}}$ and $\psi_{\rho}^{+\dot{2}}$ by complex conjugation. Analogous relations hold for the antiholomorphic fermions. Also, in the twisted sector the current operators become
\begin{subequations}
\begin{align}
 \label{eq:J3rho}
J^3 &= - \frac{1}{2}  \sum_{\rho=0}^{k-1} \psi_{\rho}^{+\dot{A}} \psi_{\rho}^{-\dot{B}} \epsilon_{\dot{A}\dot{B}},\\
J^+ &= \frac{1}{2} \left(\psi_{\rho=0}^{+\dot{A}} \psi_{\rho=0}^{+\dot{B}} \epsilon_{\dot{A}\dot{B}} + \sum_{\rho=1}^{k-1} \psi_{\rho}^{+\dot{A}} \psi_{k-\rho}^{+\dot{B}} \epsilon_{\dot{A}\dot{B}}\right),\\
J^- &= \frac{1}{2} \left(\psi_{\rho=0}^{-\dot{A}} \psi_{\rho=0}^{-\dot{B}} \epsilon_{\dot{A}\dot{B}} + \sum_{\rho=1}^{k-1} \psi_{\rho}^{-\dot{A}} \psi_{k-\rho}^{-\dot{B}} \epsilon_{\dot{A}\dot{B}}\right).
\end{align}
\end{subequations}
Using \eqref{eq:fermionsrho} the light operators are rewritten as
\begin{equation}
 \label{eq:lightoptwist}
\sum_{r=1}^k O^{++}_{(r)} =  \sum_{\rho=0}^{k-1} O^{++}_\rho,\qquad O^{++}_\rho \equiv -\frac{\ii}{\sqrt{2}} \psi^{+\dot{A}}_{\rho} \epsilon_{\dot{A} \dot{B}} \tilde{\psi}^{+\dot{B}}_{\rho},
\end{equation}
where $O^{--}$ is the complex conjugate of this. Our choice for the heavy operators in the $s=pk$ case is
\begin{equation}
 \label{eq:heavyoptwist}
S_{k,pk,\rho}^{\dot{1}} = e^{\ii \left( - \frac{\rho}{k}+\frac{1}{2}+\frac{s}{k}\right) H_\rho},\qquad S_{k,pk,\rho}^{\dot{2}} = e^{\ii \left( - \frac{\rho}{k}+\frac{1}{2}+\frac{s}{k}\right) K_\rho},
\end{equation}
with the right part given by analogous definitions with $s = 0$. The states generated by these operators are
\begin{equation}
\begin{aligned}
|s,k\rangle &\equiv \left[\Big(J^+_{-2s/k} \ldots J^+_{-2/k}\Big) \lim_{z,\bar z\to 0} \Sigma_k\, \tilde{\Sigma}_k  \otimes_{\rho=0}^{k-1}\ S^{\dot{1}}_{k,\rho} S^{\dot{2}}_{k,\rho} \tilde{S}^{\dot{1}}_{k,\rho} \tilde{S}^{\dot{2}}_{k,\rho}\right]^{N/k}|0\rangle \,.
\end{aligned}
\end{equation}
Following the same logic as in the untwisted sector, the correlator is given in terms of functions
\begin{align}
 \label{eq:corrcomptwist}
F_{pk, k, \rho}^{\dot{A}\dot{C}}(z_i) &\equiv \langle e^{\ii \left( - \frac{\rho}{k}+\frac{1}{2}+p\right) \left( H_\rho(z_1) + K_\rho(z_1)\right)} e^{-\ii \left( - \frac{\rho}{k}+\frac{1}{2}+p\right) \left( H_\rho(z_2) + K(z_2)_\rho\right)} \psi_{\rho}^{+\dot{A}}(z_3) \psi_{\rho}^{-\dot{C}}(z_4)\rangle \\ \nonumber
\times  & \prod_{\rho^\prime \neq \rho} \langle e^{\ii \left( -\frac{\rho^\prime}{k} + \frac{1}{2} + p\right)\left( H_{\rho^\prime}(z_1) + K_{\rho^\prime}(z_1)\right)} e^{-\ii \left( -\frac{\rho^\prime}{k} + \frac{1}{2} +p\right)\left( H_{\rho^\prime}(z_2) + K_{\rho^\prime}(z_2)\right)} \rangle\ \langle \Sigma_k(z_1) \Sigma_k(z_2)\rangle
\end{align}
as
\begin{equation}
 \label{eq:corrtwist}
\langle O_H(z_1) \bar{O}_H(z_2) O_L(z_3) \bar{O}_L(z_4) \rangle = \frac{1}{k} \sum_{\rho=0}^{k-1}  \frac{1}{2}\ F_{pk, k, \rho}^{\dot{A}\dot{C}}(z_i) F_{0, k, \rho}^{\dot{B}\dot{D}}(\bar{z}_i)\ \epsilon_{\dot{A}\dot{B}} \epsilon_{\dot{C}\dot{D}},
\end{equation}
where the $1/k$ factor takes care of the fact that we have the same contribution for each length $k$ strand (it would be $N/k$, but $N$ cancels out because of the normalization of the light operators). As in the untwisted sector, $F_{s, k, \rho}^{\dot{A}\dot{C}}(z_i)$ is nonzero only if $(\dot{A}, \dot{C})$ take values $(\dot{1}, \dot{2})$ or $(\dot{2}, \dot{1})$, and we have
\begin{equation}
 \label{eq:F12twist}
 \begin{aligned}
F_{pk,k,\rho}^{\dot{1}\dot{2}}(z_i) = - \frac{z_{13}^{-\frac{\rho}{k}+\frac{1}{2}+p} z_{24}^{-\frac{\rho}{k}+\frac{1}{2}+p}}{z_{12}^{2h} z_{14}^{-\frac{\rho}{k}+\frac{1}{2}+p} z_{23}^{-\frac{\rho}{k}+\frac{1}{2}+p} z_{34}}=- \frac{1}{z_{12}^{2h} z_{34}}\,z^{\frac{\rho}{k}-\frac{1}{2}-p}\,,
\end{aligned}
\end{equation}
where $h = \frac{k}{4} + \frac{s(s+1)}{k}$, $F_{s,k,\rho}^{\dot{2}\dot{1}}(z_i) = - F_{s,k,\rho}^{\dot{1}\dot{2}}(z_i)$ and $z$ is defined in \eqref{eq:zdef}. The antiholomorphic part is again obtained taking the holomorphic one, setting $s = 0$ (i.e. $p=0$) and replacing $z_i\to \bar{z}_i$ and $h\to\bar{h} = k/4$. Putting everything together we get
\begin{align}
 \label{eq:resulttwist}
\langle O_H(z_1) \bar{O}_H(z_2) O_L(z_3) \bar{O}_L(z_4) \rangle &= \frac{1/k}{z_{12}^{2h_H} \bar{z}_{12}^{2\bar{h}_H} |z_{34}|^2}\,\frac{z^{-p}}{|z|}\ \frac{1 - \left| z\right|^2}{1 - \left| z\right|^{\frac{2}{k}}}.
\end{align}
The heavy operator in the $s=pk-1$ case is obtained from the one in the $s=pk$ case by acting on it with $J^-_{2p}$ (the mode $2p$). This only changes the operator in the $\rho=0$ sector: it has the form \eqref{eq:heavyk-1}, where the action on the $\rho=0$ part of the left sector is
\begin{equation}
\label{eq:Shatop}
\hat{S}^{\dot{1}}_{k,0} = e^{\ii \left(-\frac{1}{2}+p\right) H_{\rho=0}},\qquad \hat{S}^{\dot{2}}_{k,0} = e^{\ii \left(-\frac{1}{2}+p\right) K_{\rho=0}}\;.
\end{equation}
With the same procedure as before, we have
\begin{align}
\label{eq:a.29}
F^{\dot{A}\dot{C}}_{pk-1,k,0}(z_i) &= \langle e^{\ii\left(-\frac{1}{2}+p\right)(H_0(z_1)+K_0(z_1))} e^{-\ii\left(-\frac{1}{2}+p\right)(H_0(z_2)+K_0(z_2))} \psi_0^{+\dot{A}}(z_3) \psi_0^{-\dot{C}}(z_4)\rangle \\ \nonumber
\times & \prod_{\rho^\prime=1}^{k-1} \langle e^{\ii \left( -\frac{\rho^\prime}{k} + \frac{1}{2} + p\right)\left( H_{\rho^\prime}(z_1) + K_{\rho^\prime}(z_1)\right)} e^{-\ii \left( -\frac{\rho^\prime}{k} + \frac{1}{2} +p\right)\left( H_{\rho^\prime}(z_2) + K_{\rho^\prime}(z_2)\right)} \rangle\ \langle \Sigma_k(z_1) \Sigma_k(z_2)\rangle,
\end{align}
while for $\rho\neq0$ (and in the whole right sector) we have the same functions as in \eqref{eq:corrcomptwist}, i.e. $F^{\dot{A}\dot{C}}_{pk-1,k,\rho\neq0} = F^{\dot{A}\dot{C}}_{pk,k,\rho\neq0}$. The correlator takes again the form \eqref{eq:corrtwist} and the only new object to compute is
\begin{equation}
\begin{aligned}
F_{pk-1,k,0}^{\dot{1}\dot{2}}(z_i) = - \frac{z_{13}^{-\frac{1}{2}+p} z_{24}^{-\frac{1}{2}+p}}{z_{12}^{2h} z_{14}^{-\frac{1}{2}+p} z_{23}^{-\frac{1}{2}+p} z_{34}}=- \frac{1}{z_{12}^{2h} z_{34}}\,z^{\frac{1}{2}-p},
\end{aligned}
\end{equation}
where again $h =\frac{k}{4} + \frac{s(s+1)}{k}$ and $F_{pk-1,k,0}^{\dot{2}\dot{1}}(z_i) = -F_{pk-1,k,0}^{\dot{1}\dot{2}}(z_i)$. The full correlator in the $s=pk-1$ case reads
\begin{equation}
\begin{aligned}
\langle O_H(z_1) \bar{O}_H(z_2) O_L(z_3) \bar{O}_L(z_4) \rangle &= \frac{1/k}{z_{12}^{2h_H} \bar{z}_{12}^{2\bar{h}_H} |z_{34}|^2}\,z^{-p}  \left( \left(\frac{z}{\bar{z}}\right)^{\frac{1}{2}}  + \frac{1}{|z|}\,\frac{|z|^{\frac{2}{k}} - |z|^2}{1- |z|^{\frac{2}{k}}} \right).
\end{aligned}
\end{equation}
The first correlator considered in the twisted sector corresponds to choosing $s=0$, while the second and the third correspond respectively to the $s=pk$ and the $s=pk-1$ case.

\section{Wave equation in AdS$_3/\mathbb{Z}_k$}
\label{sec:waveequation}
In this section we solve the wave equation \eqref{eq:3Deqs} for a scalar field of dimension 1, in the geometry written in \eqref{eq:6Dmetric}, with the monodromy \eqref{eq:monodromy} and the boundary condition \eqref{eq:sol1}. We will follow a route similar to the one employed in \cite{Skenderis:2008dg,Aref'eva:2016pew}, and our result generalises the one obtained in the previous works to the case with non-trivial monodromy ($\hat s\not=0)$. The boundary CFT lives on the cylinder and to induce the appropriate geometry on the boundary we will work in global AdS coordinates; we will keep careful track of the periodicity of the spatial circle, which is crucial to distinguish geometries with different values of the conical defect and to properly implement the monodromy condition. More general discussions about the dynamics of a scalar field in Lorentzian AdS of general conformal dimension, the interpretation of the normalizable modes solution, and the difference between different choice of patch can be found in \cite{Balasubramanian:1998sn}.

The AdS part of the geometry in \eqref{eq:AdSmetric} can be simplified by the redefinitions:
\begin{equation}
\label{eq:change}
t=k\,\frac{\sqrt{Q_1Q_5}}{a}\tau\quad\quad y=k\,\frac{\sqrt{Q_1Q_5}}{a}\sigma\,,\quad r=\frac{a}{k}\tan\rho\,,
\end{equation}
where the new coordinates $\tau$, $\sigma$, $\rho$ have the following domains
\begin{equation}
\rho\in \left [0,\frac{\pi}{2} \right ]\,,\quad\sigma\in \left [0,\frac{2\pi}{k} \right]\,,\quad \tau \in [0,+\infty)\,.
\end{equation}
After this change the metric takes the form 
\begin{equation}
\label{eq:metric}
ds^2_{AdS_3}=\frac{1}{\cos^2\rho}\left(-d\tau^2+d\rho^2+\sin^2\!\rho \,d\sigma^2\right)
\end{equation}
with the boundary located at $\rho=\frac{\pi}{2}$. \\
The most general solution with the prescribed monodromies involves an arbitrary sum over Fourier modes:
\begin{equation}
\label{eq:fourier}
B(\tau,\sigma,\rho)=\frac{1}{(2\pi)^2}\,e^{\ii \hat s \sigma}\sum_{l\in\mathbb{Z}}\int d\omega\, e^{\ii\omega\tau}e^{\ii lk\sigma}g(l,\omega)\chi_{l,\omega}(\rho)\,,
\end{equation}
where the choice of the function $g(l,\omega)$ encodes a particular boundary data and we assume $0\le \hat s < k$. Substituting into the wave equation we obtain a differential equation for $\chi_{l,\omega}(\rho)$ that reads
\begin{equation}
\label{eq:equation}
\chi''_{l,\omega}(\rho)+\csc\rho\sec\rho\chi'_{l,\omega}(\rho)+\left(\omega^2-(l k +\hat s)^2\csc^2\rho+\ell (\ell-2)\right)\chi_{l,\omega}(\rho)=0\,.
\end{equation}
This is an hypergeometric equation, as it is made evident by the change $x=\sin^2\rho$:
\begin{equation}
\chi_{l,\omega}''(x)+\frac{1}{x}\chi_{l,\omega}'(x)+\frac{1}{4}\left(\frac{\omega^2}{x(1-x)}-\frac{(l k +\hat s)^2}{x^2(1-x)}+\frac{1}{x(1-x)^2} \right)\chi_{l,\omega}(x)=0\,.
\end{equation}
The solution that is finite everywhere in the bulk\footnote{The form of the other independent solution can be found, for example, in \cite{Balasubramanian:1998sn}. It can be shown to contain divergences for  $x\to 0$ (i.e. $r\to 0$).} is
\begin{equation}
\label{eq:solution}
\chi_{l,\omega}(x)=x^{\frac{|l k +\hat s|}{2}}(1-x)^{\frac{1}{2}}\; {}_2 F_{1}\left(\frac{1}{2}(1+ |l k +\hat s|-\omega), \frac{1}{2}(1+|l k+\hat s|+\omega), 1+|l k +\hat s|, x  \right)\,.
\end{equation}
From the expansion of this solution near the boundary ($x=1$) one can extract the non-normalizable and the normalizable modes
\begin{equation}
\label{eq:expansion}
\begin{split}
&\chi_{l,\omega}(x)\approx \frac{\Gamma (1+|lk+\hat s|)}{\Gamma (\frac{1}{2}(1+|lk+\hat s|-\omega))\Gamma (\frac{1}{2}(1+|lk+\hat s|+\omega))}\times\\
&\Bigl\{\left[2\gamma_E+\psi (\frac{1}{2}(1+|lk+\hat s|-\omega))+\psi (\frac{1}{2}(1+|lk+\hat s|+\omega))\right](1-x)^{\frac{1}{2}}\\
&+\left[\log (1-x)\right](1-x)^{\frac{1}{2}}\Bigr\}\,,\\
\end{split}
\end{equation}
with the digamma function defined as $\psi(z)\equiv\frac{d}{dz}\log (\Gamma (z)) $, and $\gamma_E$ the Euler constant. The non-normalizable mode (the source) is the coefficient of the $\left[\log (1-x)\right](1-x)^{\frac{1}{2}}$ term and the normalizable mode (the VEV) is the term proportional to  $(1-x)^{\frac{1}{2}}$. Reverting to the original coordinates, these two terms correspond to the ones shown in \eqref{eq:sol1}. A delta function source at the boundary is obtained by tuning the function $g(l,\omega)$ in \eqref{eq:fourier} in such a way that the non-normalizable term has constant Fourier transform; this is achieved setting
\begin{equation}
\label{eq:coeff}
\begin{split}
g(l,\omega)=\frac{\Gamma (\frac{1}{2}(1+|lk+\hat s|-\omega))\Gamma (\frac{1}{2}(1+|lk+\hat s|+\omega))}{\Gamma (1+|lk+\hat s|)}\,.
\end{split}
\end{equation}
The coefficient of the normalizable term, denoted as $b_1(\tau,\sigma)$ in \eqref{eq:sol1}, is then found from \eqref{eq:expansion} to be
\begin{equation}
\label{eq:sum}
b_1(\tau,\sigma)=\sum_{l\in\mathbb{Z}}\int \frac{d\omega}{(2\pi)^2} \,e^{\ii\omega\tau + \ii(lk+\hat s)\sigma}\Bigl[\psi (\frac{1}{2}(1+|lk+\hat s|-\omega))+\psi (\frac{1}{2}(1+|lk+\hat s|+\omega))  +2\gamma_E\Bigr]\,.
\end{equation}
In order to perform the sum we use the series representation of the digamma function
\begin{equation}
\psi(z)=-\gamma_E+\sum_{n=0}^{\infty}\left(\frac{1}{n+1}-\frac{1}{n+z}\right)\,.
\end{equation}
Separating the term with $l=0$ in the sum, and forgetting contact terms coming from summation over constants Fourier modes we have
\begin{equation}
\begin{split}
b_1(\tau,\sigma)&=\!\sum_{n=0}^{\infty}\Biggl[\sum_{l=0}^{\infty}\int\!\! \frac{d\omega}{(2\pi)^2}\,e^{\ii\omega\tau + \ii(lk+\hat s)\sigma}\!\left(\frac{2}{\omega-(l k +\hat s)-1-2n}-\frac{2}{\omega+(l k +\hat s)+1+2 n} \right)\\
&+\sum_{l=1}^{\infty}\int \frac{d\omega}{(2\pi)^2}\,e^{\ii\omega\tau - \ii(lk-\hat s)\sigma}\left(\frac{2}{\omega-(l k -\hat s)-1-2n}-\frac{2}{\omega+(l k -\hat s)+1+2 n} \right)\Biggr]\,.
\end{split}
\end{equation}
As usual, to define the $\omega$-integral one has to pick the integration contour: we choose the Feynman prescription, which allows the Wick rotation to Euclidean space and hence comparison with the CFT correlator, which is evaluated on the Euclidean plane. The integral is thus readily computed and yields
\begin{equation}
\begin{split}
b_1(\tau,\sigma)&=-\frac{\ii}{2\pi}\sum_{n=0}^{\infty}\Biggl[\sum_{l=0}^{\infty} e^{\ii (l k +\hat s)\sigma} e^{-\ii(lk +\hat s +1 +2n)\tau}+\sum_{l=1}^{\infty} e^{-\ii (l k -\hat s)\sigma} e^{-\ii(lk -\hat s +1 +2n)\tau}\Biggr]\\
&=-\frac{\ii}{2\pi}\frac{e^{\ii \hat s \sigma}}{e^{\ii \tau}-e^{-\ii\tau}} \left[\frac{e^{-\ii \hat s\tau}}{1-e^{\ii k (\sigma-\tau)}}+\frac{e^{\ii \hat s \tau}}{e^{\ii k (\sigma+\tau)}-1}\right]\,.
\end{split}
\end{equation}
Re-expressing the result in the original physical coordinates defined in \eqref{eq:change}, and suppressing the overall numerical coefficient (which is not meaningful as we did not keep  track of the normalization of the operators), we finally obtain
\begin{equation}
\begin{split}
b_1(t,y) &=-\ii \frac{e^{\ii \hat s \frac{y}{R\,k}}}{e^{\ii \frac{t}{R\,k}} - e^{-\ii \frac{t}{R\,k}}}\,\left[\frac{e^{\ii \frac{t-y}{R}}}{e^{\ii \frac{t-y}{R}}-1}e^{-\ii \hat s\frac{t}{R\,k}}+\frac{1}{e^{\ii \frac{t+y}{R}}-1} e^{\ii \hat s \,\frac{t}{R\,k}}\right]\\
&=-\ii \left(\frac{z}{\bar z}\right)^{\frac{\hat s}{2k}}\frac{1}{|z|^{\frac{1}{k}}-|z|^{-\frac{1}{k}}}\,\left[\frac{\bar z}{\bar z -1} |z|^{-\frac{\hat s}{k}}+\frac{1}{z-1} |z|^{\frac{\hat s}{k}}\right]\,.
\end{split}
\end{equation}

\providecommand{\href}[2]{#2}\begingroup\raggedright\endgroup

 \end{document}